\documentclass[journal=nalefd,manuscript=letter]{achemso}

\usepackage[T1]{fontenc}
\usepackage[latin1]{inputenc}
\usepackage{lmodern}
\usepackage{amsmath}
\usepackage{amssymb}
\usepackage{pst-all}
\usepackage{psfrag}
\usepackage{graphicx}
\usepackage{dcolumn}
\usepackage{bm}
\usepackage{epsfig}

    \def\db{{\bf d}}

\def\Fb{{\bf F}}    \def\Eb{{\bf E}}

\title{Graphene plasmonics: A platform for strong light-matter interaction}

\author{Frank~H.~L.~Koppens}
\affiliation[ICFO]{ICFO-Institut de Ci\'encies Fot\'oniques, Mediterranean Technology Park, 08860 Castelldefels (Barcelona), Spain}
\email{frank.koppens@icfo.es}
\author{Darrick~E.~Chang}
\affiliation[Caltech]{California Institute of Technology, Pasadena, California 91125, USA}
\author{F.~Javier~Garc\'{\i}a~de~Abajo}
\affiliation[IO-CSIC]{Instituto de \'Optica - CSIC, Serrano 121, 28006 Madrid, Spain}
\affiliation[ORC-UoS]{Optoelectronics Research Centre, University of Southampton, Southampton SO17 1BJ, UK}
\email{J.G.deAbajo@csic.es}

\begin{document}
\begin{abstract}
Graphene plasmons provide a suitable alternative to noble-metal plasmons because they exhibit much larger confinement and relatively long propagation distances, with the advantage of being highly tunable via electrostatic gating. We report strong light-matter interaction assisted by graphene plasmons, and in particular, we predict unprecedented high decay rates of quantum emitters in the proximity of a carbon sheet, large vacuum Rabi splitting and Purcell factors, and extinction cross sections exceeding the geometrical area in graphene ribbons and nanometer-sized disks. Our results provide the basis for the emerging and potentially far-reaching field of graphene plasmonics, offering an ideal platform for cavity quantum electrodynamics and supporting the possibility of single-molecule, single-plasmon devices.
\end{abstract}

Surfaces plasmons (SPs), the electromagnetic waves coupled to charge excitations at the surface of a metal, are the pillar stones of applications as varied as ultrasensitive optical biosensing\cite{KWK97,TJO05,paper168}, photonic metamaterials\cite{Z11}, light harvesting\cite{P08_2,AP10}, optical nano-antennas\cite{NV11}, and quantum information processing\cite{CSH06,DSF10,SSR10,GMM11}. However, even noble metals, which are widely regarded as the best available plasmonic materials\cite{WIN10}, are hardly tunable and exhibit large ohmic losses that limit their applicability to optical processing devices.

In this context, doped graphene emerges as an alternative, unique two-dimensional plasmonic material that displays a wide range of extraordinary properties\cite{JBS09}. This atomically thick sheet of carbon is generating tremendous interest due to its superior electronic and mechanical properties\cite{NGM04,ZTS05,BSL06,GN07,NGM05,CGP09,G09}, which originate in part from its charge carriers of zero effective mass (the so-called Dirac fermions\cite{NGM05}) that can travel for micrometers without scattering, even at room temperature\cite{BSJ08}. Furthermore, rapid progress in growth and transfer techniques have sparked expectations for large-scale production of graphene-based devices and a wide range of potential applications such as high-frequency nanoelectronics, nanomechanics, transparent electrodes, and composite materials\cite{GN07}.

Recently, graphene has also been recognized as a versatile optical material for novel photonic\cite{VE11} and optoelectronic applications\cite{BSH10}, such as solar cells, photodetectors\cite{XML09}, light emitting devices, ultrafast lasers, optical sensing\cite{SLL10}, and metamaterials\cite{PLS10}. The outstanding potential of this atomic monolayer is emphasized by its remarkably high absorption\cite{NBG08,MSW08} $\approx\pi\alpha\approx2.3$\%, where $\alpha=e^2/\hbar c\approx1/137$ is the fine-structure constant. Moreover, the linear dispersion of the Dirac fermions enables broadband applications, in which electric gating can be used to induce dramatic changes in the optical properties\cite{LHJ08}.

All of these photonic and optoelectronic applications rely on the interaction of propagating far-field photons with graphene. Additionally, SPs bound to the surface of doped graphene exhibit a number of favorable properties that make graphene an attractive alternative to traditional metal plasmonics. In particular, graphene plasmons are confined to volumes $\sim10^6$ times smaller than the diffraction limit, thus facilitating strong light-matter interactions. Furthermore, dramatic tuning of the plasmon spectrum is possible through electrical or chemical modification of the charge carrier densities\cite{MSW08,CPB11}, and a Fermi energy as high as $E_F=$1-2\,eV has been recently realized\cite{CPB11,EK10}. And lastly, the electronic structure of graphene and the ability to fabricate large, highly crystalline samples should give rise to SP lifetimes reaching hundreds of optical cycles, thereby circumventing one of the major bottlenecks facing noble-metal plasmonics.

Here, we show that these properties can be used to tailor extremely strong light-matter interactions at the quantum level. In particular, we consider the interaction between a single quantum emitter and single SPs in graphene, and show that the extreme mode confinement yields ultrafast and efficient decay of the emitter into single SPs of a proximal, doped graphene sheet (see Fig.\ 1a). More precisely, we analyze confinement in 2D (homogeneous graphene), 1D (nanoribbon), and 0D (nanodisk) geometries. We find an increased degree of field enhancement and interaction strengths with reduced dimensionality, ultimately yielding in 0D structures decay rates exceeding the natural decay rate by six orders of magnitude. Consequently, graphene opens up a novel route to {\it quantum plasmonics} and quantum devices that have so far been difficult to achieve with conventional metal plasmonics. Beyond the controlled enhancement and channeling of emission, graphene should also assist the exploration of fundamentally new regimes of quantum plasmonic interactions at the nanoscale. Specifically, we predict observable vacuum Rabi splittings in our proposed nanostructures, enabling a single SP to be emitted and then re-absorbed. Finally, we show that these structures have resonant extinction cross-sections greatly exceeding their geometrical cross-sections, despite the small volume occupied by this thin material, thus rendering the effects observable in practice and paving the way to advanced optoelectronic applications in which photon absorption is dramatically enhanced.

\subsection*{Optical response of graphene}

The photonic properties of this material can be fully traced back to its in-plane conductivity $\sigma(k_\parallel,\omega)$, which is in general a function of parallel wave vector $k_\parallel$ and frequency $\omega$. This quantity is mainly controlled by electron-hole pair (e-h) excitations that can be divided into intraband and interband contributions (see Fig.\ 1b). Within the random-phase approximation\cite{WSS06,HD07} (RPA), the conductivity of graphene in the local limit ($k_{\parallel}\rightarrow0$) reduces to
\begin{equation}
\sigma(\omega)=\frac{e^2E_F}{\pi\hbar^2}\frac{i}{\omega+i\tau^{-1}}
+\frac{e^2}{4\hbar}\left[\theta(\hbar\omega-2E_F)+\frac{i}{\pi}
\log\left|\frac{\hbar\omega-2E_F}{\hbar\omega+2E_F}\right|
\right] \label{Falkovsky}
\end{equation}
(we take $E_F>0$, see Fig.\ 1c). The first term of Eq.\ (\ref{Falkovsky}) describes a Drude model response for intraband processes, conveniently corrected to include a finite relaxation time $\tau$, for which we use a conservative value ($\tau\approx10^{-13}\,$s at $E_F=0.1\,$eV$^{-1}$) extracted from the measured, impurity-limited DC mobility\cite{NGM04,NGM05} (see Appendix). The remaining terms arise from interband transitions, which produce significant losses at energies near and above $2E_F$ (i.e., ${\rm Re}\{\sigma\}=e^2/4\hbar$, resulting in the well-established 2.3\% absorption). Equation\ (\ref{Falkovsky}) is valid at zero temperature, but we actually employ a finite-temperature extension\cite{FV07} at $T=300\,$K (see Appendix), in which the step function is smeared out by thermal effects (Fig.\ 1c). The local limit (Eq. (\ref{Falkovsky})) produces decay rates in reasonable agreement with the nonlocal RPA (see Appendix), except in the region near the interband transition onset or for distances below $v_F/\omega$. 

\subsection*{Plasmons in homogeneous graphene: Extraordinary confinement}

For sufficiently high doping ($E_F>\omega$) graphene can sustain $p$-polarized SPs propagating along the sheet with wave vector $k_{\rm sp}\approx i(\epsilon+1)\omega/4\pi\sigma$ and electric field profile $\Eb\sim\exp[k_{\rm sp}(ix-|z|)]$. These electrostatic expressions, valid for $\hbar\omega\gg\alpha E_F$ (see Appendix), reduce upon insertion of the Drude formula (first term of Eq.\ (\ref{Falkovsky})) to
\begin{equation}
k_{\rm sp}\approx(\hbar^2/4e^2E_F)\,(\epsilon+1)\,\omega(\omega+i/\tau),
\label{kspDrude}
\end{equation}
clearly showing a quadratic dependence of $k_{\rm sp}$ on $\omega$, which is characteristic of 2D electron gases\cite{S1967_2} (see Fig.\ 1e).

The remarkable degree of confinement provided by the graphene is clear from the ratio of SP to free-space-light wavelengths $\lambda_{\rm sp}/\lambda_0\approx[4\alpha/(\epsilon+1)](E_F/\hbar\omega)$ derived from Eq.\ (\ref{kspDrude}). In addition, the out-of-plane wave vector $\sim ik_{\rm sp}$ indicates an equally tight confinement to dimensions $\sim\lambda_{\rm sp}/2\pi$ in the transverse direction $z$.

Interestingly, the in-plane propagation distance ($1/e$ decay in amplitude), given by $1/{\rm Im}\{k_{\rm sp}\}$, reaches values well above 100 SP wavelengths (see inset of Fig.\ 1e) and drops rapidly at high energies when the plasmon has sufficient energy to generate e-h's and the dispersion relation enters the intraband region (see Fig.\ 1b). Figure\ 1e shows clearly that for increasing $E_F$ the plasmons become narrower because the damping rate $1/\tau$ decreases relative to the photon frequency $\omega$ (Fig.\ 1e).

\subsection*{Strong SP-emitter coupling in graphene}

Our first realization is that when an emitter such as an excited molecule or a quantum dot is placed close to doped homogeneous graphene, the emission rate is enhanced by 1-5 orders of magnitude and the emitted energy is mainly converted into a plasmon in the carbon sheet, as illustrated in the near-electric-field-amplitude plot of Fig.\ 1a.

The decay rate $\Gamma$ is proportional to the strength of the coupling between the transition dipole matrix element $\db$ and the electromagnetic modes acting on it, including the plasmon. This can be related to the electric field induced by the dipole on itself $\Eb^{\rm ind}$ (i.e., the field reflected by the graphene) as\cite{NH06}
\begin{equation}
\Gamma=\Gamma_0+\frac{2}{\hbar}{\rm Im}\{\db^*\cdot\Eb^{\rm ind}\}, \label{GamGam}
\end{equation}
where $\Gamma_0=4k_0^3|\db|^2/3\hbar$ is the free-space decay rate. For homogeneous graphene, the induced field is related to the Fresnel coefficients. This yields an exact relation for $\Gamma$ as an integral over parallel wave vector contributions\cite{NH06} (see Appendix), which we use in the calculations presented in Figs.\ 1,2. However, it is instructive to explore the electrostatic limit, which is accurate for small distances compared to the emission wavelength:
\begin{equation}
\Gamma\approx\Gamma_0+\frac{2}{\hbar}\left(|{\bf d}_\parallel|^2+2|{\bf d}_\perp|^2\right)\int_0^\infty k_\parallel^2\,dk_\parallel\,{\rm Im}\left\{\frac{-1}{\epsilon+1+4\pi ik_\parallel\sigma/\omega}\right\}\,e^{-2k_\parallel z}, \label{GammaNR}
\end{equation}
where $z$ is the emitter-graphene separation and $\parallel$ ($\perp$) denotes components parallel (perpendicular) to the graphene. The exponential in the above integral effectively suppresses the contribution of wave vectors $k_\parallel\gg1/z$.

The spectral dependence of the decay rate is represented in Fig.\ 1d (solid curves) for various values of $E_F$ when the emitter is placed 10\,nm away from graphene supported on silica ($\epsilon=2$). The rate is peaked at a photon energy below $E_F$, before dropping dramatically and finally converging to a common $E_F$-independent value at energies above $2E_F$. This behavior can be understood as follows. When the SP mode is well defined, the integral in Eq. (\ref{GammaNR}) separates into two distinct contributions, a sharp pole associated with emission into the SPs and a broad background out to wave vectors $k_\parallel\sim1/z$ associated with emission into lossy channels. The pole contribution yields the SP emission rate
\begin{equation}
\Gamma_{\rm sp}\approx\frac{(2\pi)^4}{(\epsilon+1)\hbar}\left(|{\bf d}_\parallel|^2+2|{\bf d}_\perp|^2\right)\frac{e^{-4\pi z/\lambda_{\rm sp}}}{\lambda_{\rm sp}^3}.
\label{Gammasp}
\end{equation}
This SP contribution (dashed curves in Fig.\ 1d) dominates the decay below $E_F$ and is responsible for the emission maximum in that region. For distances $z\ll\lambda_{\rm sp}$, the decay rate is enhanced by a factor $[3\pi f/2(\epsilon+1)]\,(\lambda_0/\lambda_{\rm sp})^3$ relative to the free-space rate, where $\lambda_0$ ($\gg\lambda_{\rm sp}$) is the light wavelength and $f=1$ ($f=2$) for parallel (perpendicular) polarization. This explains the high values obtained for $\Gamma/\Gamma_0$ in Fig.\ 1d, which can be attributed to the smaller mode volume ($(\lambda_0/\lambda_{\rm sp})^2$ factor) and the reduced group velocity (additional $\lambda_0/\lambda_{\rm sp}$ factor).

The noted spectral dip in the decay corresponds to the  onset of interband transitions at $\hbar\omega=2E_F$ and decay into lossy channels as the dominant emission mechanism (see Fig.\ 1b,c). At energies above this dip, the response is dominated by interband transitions, whereas intraband and SP excitations become unimportant, and the rate follows a common profile similar to undoped graphene ($E_F=0$). In fact, undoped graphene exhibits a novel phenomenon of strong quenching induced by the high conductivity of the carbon sheet\cite{SS08,KCK10,SKB10,CBN10}.

It is important to note that the radiative emission rate near graphene is comparable to $\Gamma_0$ and therefore negligible compared to SP launching. The carbon sheet is thus eager to absorb most of the optical energy released in its vicinity. Consequently, the emitter serves as an extremely efficient excitation source of single SPs in graphene.

A large degree of control over the emission rate can be gained by situating the emitter at different distances with respect to the graphene layer, as Fig.\ 2 illustrates. The near-field plots of Fig.\ 2b,c describe full coupling to SPs at small distance and partial coupling at larger separations. The decay and plasmon launching rates both exhibit an exponential fall-off with distance predicted by Eq.\ (\ref{Gammasp}) within the spectral range for which the SPs are well defined. At larger energies above the plasmon cutoff, one recovers the same rate as in undoped graphene, characterized by a $1/z^6$ dependence at large separations (see inset).

\subsection*{Engineering plasmonic nanostructures}

While we have thus far studied extended graphene sheets, their patterning into nanometer-sized cavities yields additional benefits such as extreme field confinement in 3D, engineering of resonances, and enhanced coupling efficiency with the far-field.

\subsection*{\underline{Nanoribbons}}

A first example of confinement along one spatial dimension is provided by nanoribbons. The results of a thorough theoretical analysis are shown in Fig.\ 3. These calculations show that nanoribbons offer an efficient means of exciting surface SPs in graphene, which can subsequently drive a nearby emitter. Figure\ 3a depicts the extinction cross section of self-standing graphene ribbons for light incident normal to the graphene plane with its polarization across the ribbon (supported ribbons lead to similar results, as shown in the Appendix). Plasmon confinement is clear from the approximate scaling of the photon energy with the inverse of the square root of the ribbon width. The cross section is quite high, demonstrating very efficient excitation of SPs. Interestingly, the cross section exceeds the graphene area in some cases (black regions). Simultaneously, due to the large wave vector mismatch between SPs and far-field, the scattering of plasmons back into photons is very weak, quantified by an elastic-scattering contribution to the cross-section that turns out to be more than two orders of magnitude smaller than the total cross-section. The latter is therefore dominated by the excitation of SPs that are ultimately dissipated in the carbon structure.

The SP modes exhibit intense field focusing near the edges. The number of induced-charge nodes coincides with the order of the mode (1-3), so that odd modes display a net dipole moment (e.g., modes 1 and 3 in Fig.\ 3d). In contrast, even modes (e.g., mode 2) couple rather inefficiently to external light because they have higher multipolar character.

When a line emitter is placed right above the center of the ribbon, odd modes can be excited with polarization parallel and across the ribbon (Fig.\ 3b), whereas even modes couple to perpendicular polarization (Fig.\ 3c). The decay rate is comparable in magnitude to that in front of homogeneous graphene (cf. Figures\ 1 and 3). However, in contrast to homogenous graphene, ribbons provide a very efficient way to drive the emitter by external illumination because confined plasmons can be excited by an incident plane wave. Effectively the cross section at the position of the emitter in Fig.\ 3b is increased by a factor $\sim750$. This reflects the ratio of the electric field intensity at the position of the emitter, situated 10\,nm away from the carbon sheet, to the incident intensity.

\subsection*{\underline{Nanodisks}}

Although ribbons offer an efficient way of exciting SPs, the decay rate of a nearby emitter is comparable to a graphene homogeneous sheet, and low photon energies are required. In order to boost light-matter interactions at higher photon energies,  confinement in all dimensions is desirable. Plasmon confinement in all directions is achieved using the circular disk cavities illustrated in Fig.\ 4. This leads to narrow resonances compared to ribbons and homogeneous graphene. For simplicity we consider self-standing disks, although like in the ribbons, supported disks lead to similar qualitative conclusions. The decay rate (Fig.\ 4a) is significantly boosted at resonance frequencies that can extend up to above $E_F$. This allows one to reach the near-infrared (NIR) region with attainable levels of doping\cite{EK10,CPB11}.

The light extinction cross-section is also peaked at the SP resonances (Fig.\ 4b), reaching values up to one order of magnitude larger than the disk area. However, similar to the other geometries, the radiative emission rate (dashed curves in Fig.\ 4a) is systematically below 1\% of the total decay rate (solid curves) in all cases under consideration. This indicates that SPs can be efficiently excited by external illumination, but once produced they stay in the graphene for up to a few hundred optical cycles (see $Q$ factors below) with negligible out-coupling to far-field radiation.

The modes emerging in the spectra of Fig.\ 4a have either $m=0$ or $m=1$ azimuthal symmetry when the emitter is situated above the center of the disk and is polarized perpendicular or parallel to the carbon sheet, respectively. The near electric-field of these modes, represented in Fig.\ 4c,d, clearly shows that the $m=1$ plasmon is dipolar, and thus couples efficiently to incident light, in contrast to the $m=0$ plasmon.

Detailed inspection of the $E_F$ and disk-size dependence of these SPs reveals the following properties (see Appendix): the scaling of the plasmon frequency is inherited from the $\omega_p\propto\sqrt{E_F/\lambda_{\rm sp}}$ scaling in homogeneous graphene, so that it increases with $\sqrt{E_F}$ and decreases with the inverse of the square root of the diameter $D$ (in particular, $\lambda_{\rm sp}\sim D,\,3D$ for $m=0,\,1$); maximum Purcell factors $\Gamma/\Gamma_0\sim10^6-10^7$ are consistently obtained; the quality factors (extracted from the peak frequency divided by the spectral FWHM) qualitatively follow the relation $Q\approx\omega_p\tau$ and reach values above 100 for $\tau\sim10^{-13}\,$s.

It should be noted that for such large predicted enhancements of decay rate, the perturbative treatment of light-matter interactions as described by Eq.\ (\ref{GamGam}) can break down, giving rise to a new regime of quantum behavior. Physically, the emitter cannot exponentially decay into the SPs at a rate faster than the plasmonic cavity linewidth $\kappa=\omega_p/Q$.  As described below, once $\Gamma_{\rm sp}>\kappa$, it should be possible to observe a vacuum Rabi splitting, indicating that an emitted SP can be reversibly and coherently re-absorbed by the emitter\cite{TRK92,YSH04}.

\subsection*{A new regime: Plasmonic vacuum Rabi splitting}

The large $Q$ factors and field concentrations in graphene disks are ideally suited to perform quantum optics down to the single-photon level. Here we take a simplified model consisting of a two-level quantum emitter (e.g., a quantum dot or a molecule) interacting with a near-resonant single mode of a graphene disk. Such a system is characterized by the Jaynes-Cummings Hamiltonian\cite{JC1963} $H=H_0+H_{\rm int}+H_{\rm ext}$, where
\begin{equation}
H_0=\hbar\omega_p\left(a^+a+\frac{1}{2}\right)+\hbar\omega_0\,\sigma^+\sigma-i\hbar\frac{\kappa}{2}a^+a-i\hbar\frac{\Gamma_0}{2}\sigma^+\sigma\nonumber
\end{equation}
is the non-interacting part, which includes the plasmon mode of energy $\hbar\omega_p$ and its creation and annihilation operators $a^+$ and $a$ (first term), as well as the unperturbed quantum emitter (second term), whose excited state of energy $\hbar\omega_0$ is created by the operator $\sigma^+=|1\rangle\langle 0|$, connecting its ground $|0\rangle$ and excited $|1\rangle$ levels. We introduce in $H_0$ non-Hermitian damping terms to account for inelastic decay channels of both the plasmon mode and the excited emitter (e.g., through relaxation and radiative emission) with rates $\kappa$ and $\Gamma_0$, respectively. This is consistent with a quantum jump formalism to describe this open quantum system\cite{MS99}. The SP-emitter interaction is contained in
\begin{equation}
H_{\rm int}=i\hbar g\left(\,a^+\sigma-a\sigma^+\right), \label{Hint}
\end{equation}
and $H_{\rm ext}$ represents the coupling with an external light field (see Appendix).

In this model, the SP-emitter coupling $g$ is determined by comparison with the above electromagnetic calculations (Eq.\ (\ref{GamGam})), which correspond to the low-coupling limit ($g\ll\kappa$); assuming $\omega_0=\omega_p$, the quantum model yields\cite{L1983} $\Gamma=\Gamma_0+4g^2/\kappa$. In our case, $\Gamma_0\ll\kappa$ and thus we find $g\approx\sqrt{\kappa\Gamma}/2$.

The properties of the Jaynes-Cummings Hamiltonian are well-studied\cite{MS99}. When the emitter and cavity are on resonance and the system is in the strong coupling regime ($g>\kappa,\,\Gamma_0$), the single-excitation dressed eigenstates consist of even and odd superpositions of the excited state of the emitter and a single photon, which can be distinctly resolved. In this regime, an initially excited emitter will undergo damped Rabi oscillations at a rate $g$, where the emitted photon can be re-absorbed before it leaves the cavity.

The ratio $g/\kappa$ reaches a maximum value $\sim4$ for 100\,nm self-standing graphene disks, assuming a reasonable value of the natural decay rate $\Gamma_0=5\times10^7\,$s$^{-1}$, as we show in Fig.\ 5a,b. Actually, $g/\kappa$ is clearly above one for a wide range of doping and disk-size parameters, which indicates that the strong-coupling regime is robust.

A simple signature of strong coupling can be observed in avoided crossings of the extinction cross-section of the combined disk-emitter system, $\sigma^{\rm ext}(\omega)=4\pi k_0\;{\rm Im}\{\alpha(\omega)\}$, where $\alpha(\omega)$ is the polarizability of the combined system (see Appendix). We show in Fig.\ 5c the extinction cross-section for an emitter of excited energy $\hbar\omega_0=0.3\,$eV and natural decay rate $\Gamma_0=5\times10^7\,$s$^{-1}$ situated 10\,nm above the center of a self-standing 100\,nm graphene disk. A pronounced vacuum Rabi splitting is observed that can be probed within a single device by changing the doping level.

\subsection*{Conclusion and outlook}

Here, we have described powerful and versatile building blocks for advanced graphene plasmonic circuits. These ideas take advantage of the unique combination of extreme field confinement, device tunability and patterning, and low losses that emerge from the remarkable structure of graphene and current experimental capabilities for fabrication. These advances are expected to both remove a number of obstacles facing traditional metal plasmonics, and facilitate new possibilities for manipulating light-matter interactions at the nanoscale down to the single-SP level. The simultaneous large bandwidths and field enhancements, for example, should enable novel low-power, ultrafast classical or quantum optical devices. The strong coupling between single emitters and single SPs could be used to construct fast quantum networks, or simulate exotic strongly-interacting condensed matter systems\cite{LS98}. Our proposed techniques could also be potentially applied to manipulate or couple together more exotic excitations in graphene, such as thermoplasmons in undoped graphene\cite{V06_2} or $s$-polarized plasmon modes\cite{MZ07}. Finally, while we have demonstrated the feasibility of graphene plasmonics via free-space excitation and detection, the possible applications should be even further enhanced with the advent of novel devices such as SP sources, detectors, lasers, optical switches and interconnects, plasmon-enhanced photodetectors, and other graphene-based nano-optical elements.

{\bf Acknowledgements} This work has been supported in part by the Spanish MICINN (MAT2010-14885 and Consolider NanoLight.es), Fundaci\'o Cellex Barcelona, and the European Commission (FP7-ICT-2009-4-248909-LIMA and FP7-ICT-2009-4-248855-N4E). D.E.C. acknowledges support from the NSF (Grant No. PHY-0803371) and the Gordon and Betty Moore Foundation through Caltech's Center for the Physics of Information.

\newpage
\begin{figure}
\centerline{\includegraphics*[width=8cm]{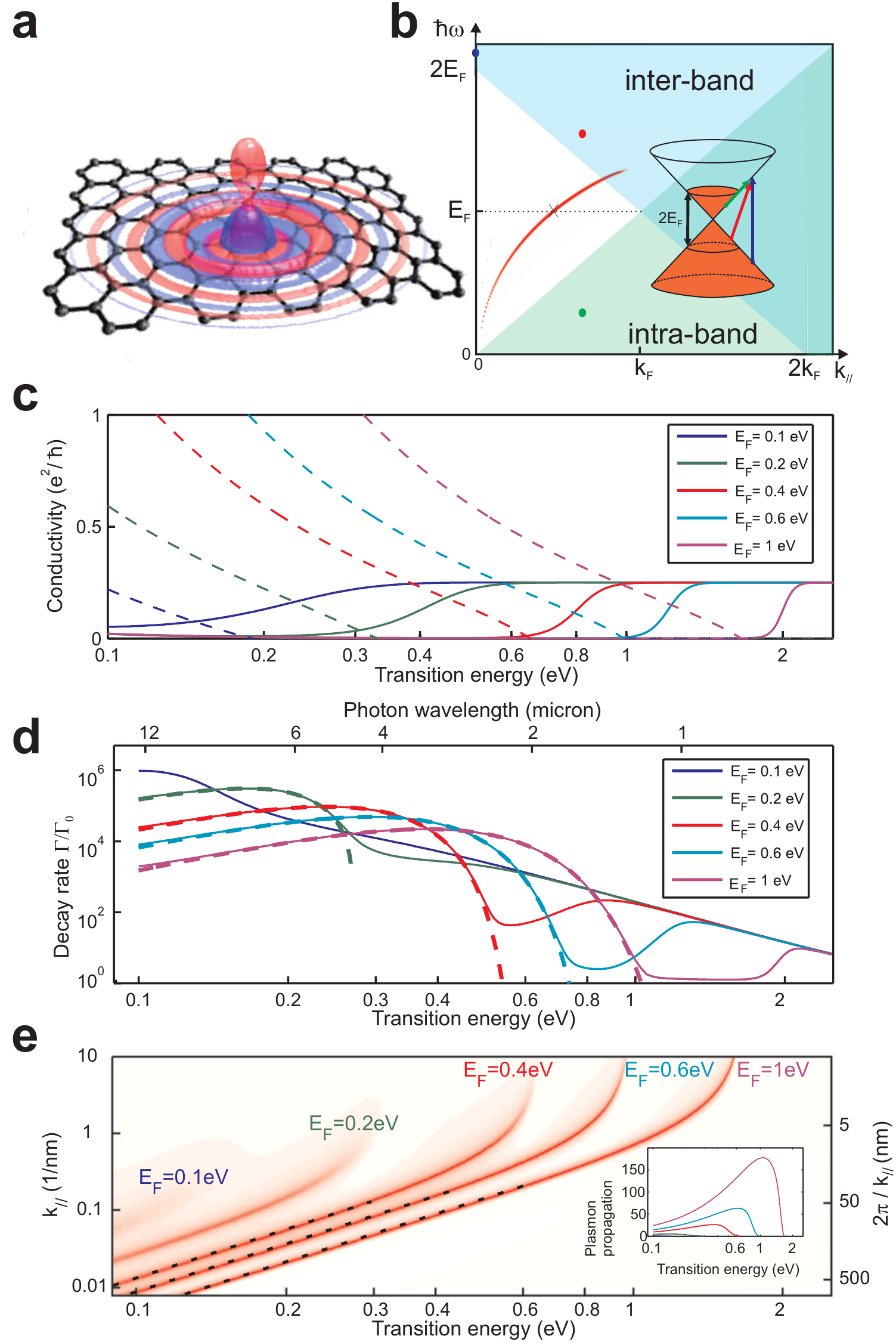}}
\caption{{\bf Coupling of a dipole emitter to doped homogeneous-graphene plasmons.}  {\bf a,} Near electric field produced by a perpendicular dipole situated 10\,nm away from doped graphene. The photon and Fermi energies ($\hbar\omega$ and $E_F$) are both 0.5\,eV. The real (imaginary) part of the perpendicular electric field is shown as a red (blue) 3D contour. {\bf b,} Optical dispersion diagram showing the surface plasmon (SP) mode for $E_F=0.5\,$eV, as well as intra- and interband transitions in graphene. {\bf c,} Real (solid curves) and imaginary (dashed curves) parts of the conductivity of doped graphene. {\bf d,} Decay rate of an excited emitter in front of doped graphene as a function of photon emission energy $E_F$. The rate $\Gamma$ is normalized to the free-space value $\Gamma_0$. The emission dipole is perpendicular to the graphene and placed 10\,nm away from it. Solid curves show the total decay rate, whereas dashed curves stand for the contribution of SP excitation. {\bf e,} Plasmon dispersion relation in doped graphene. The contour plot shows the Fresnel reflection coefficient $|r_p|$ for various values of $E_F$. The dashed lines correspond to the Drude model (Eq.\ (\ref{kspDrude})). The SP wave vector $k_{\rm sp}$ exhibits a quadratic dependence on plasmon energy. The inset shows the propagation distance $1/{\rm Im}\{k_{\rm sp}\}$ in units of the SP wavelength $\lambda_{\rm sp}=2\pi/{\rm Re}\{k_{\rm sp}\}$. The graphene is considered to lie on an $\epsilon=2$ substrate in all cases.} \label{Fig1}
\end{figure}

\begin{figure}
\centerline{\includegraphics*[width=10cm]{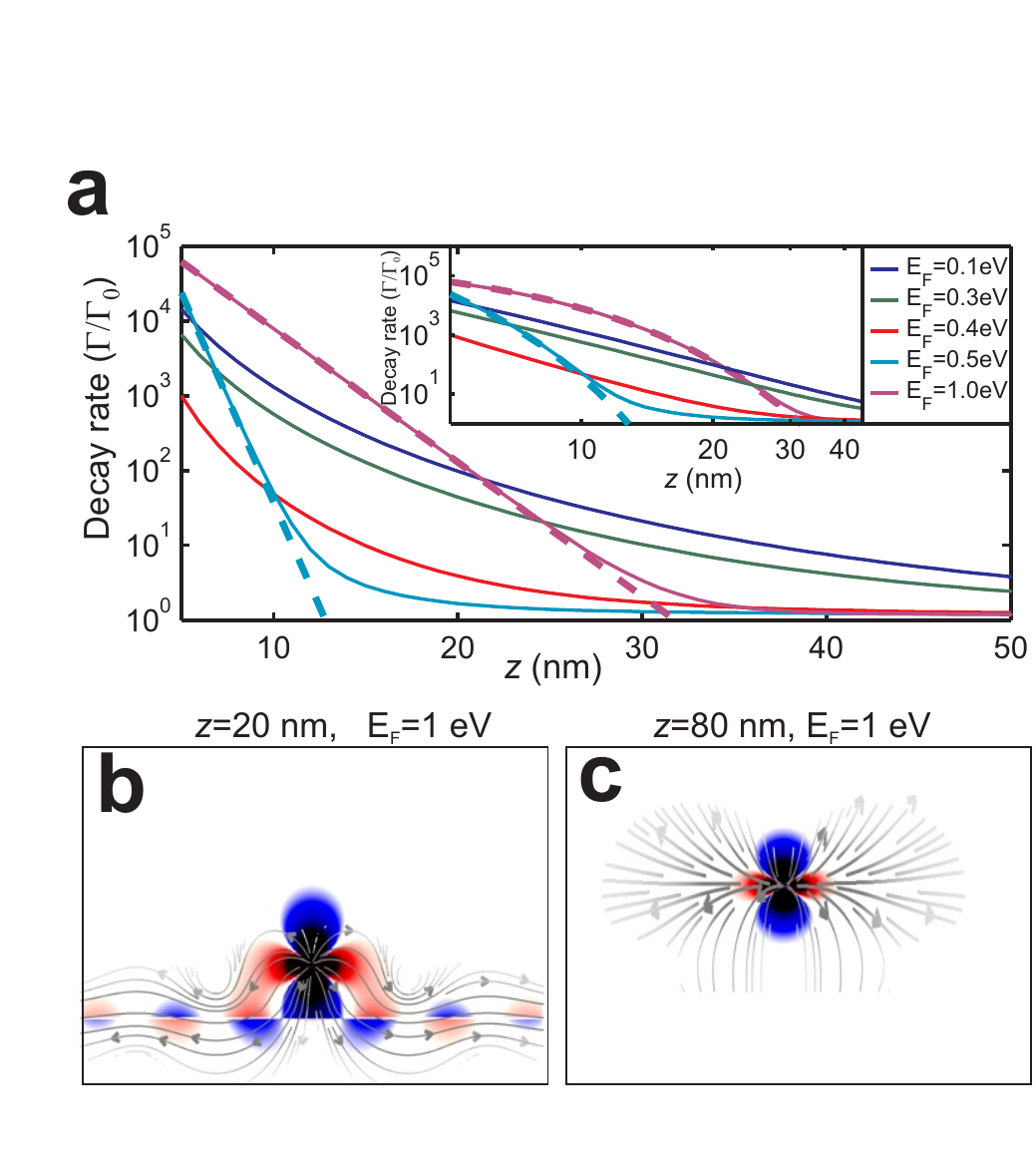}}
\caption{{\bf Distance dependence of emitter-plasmon coupling.}  {\bf a,} Variation of the decay rate with distance to the graphene for an emitter polarized perpendicular to the carbon sheet. The rate is normalized to the free-space value. Solid (dashed) curves show the total (SP-mediated) decay rate. {\bf b,c,} Near electric-field intensity for two different graphene-emitter separations and a Fermi energy $E_F=1\,$eV. Poynting vector lines are superimposed to the contour plots, with their strength shown in gray scale. The photon energy is 0.5\,eV and the substrate has $\epsilon=2$ in all cases.} \label{Fig2}
\end{figure}

\begin{figure}
\centerline{\includegraphics*[width=15cm]{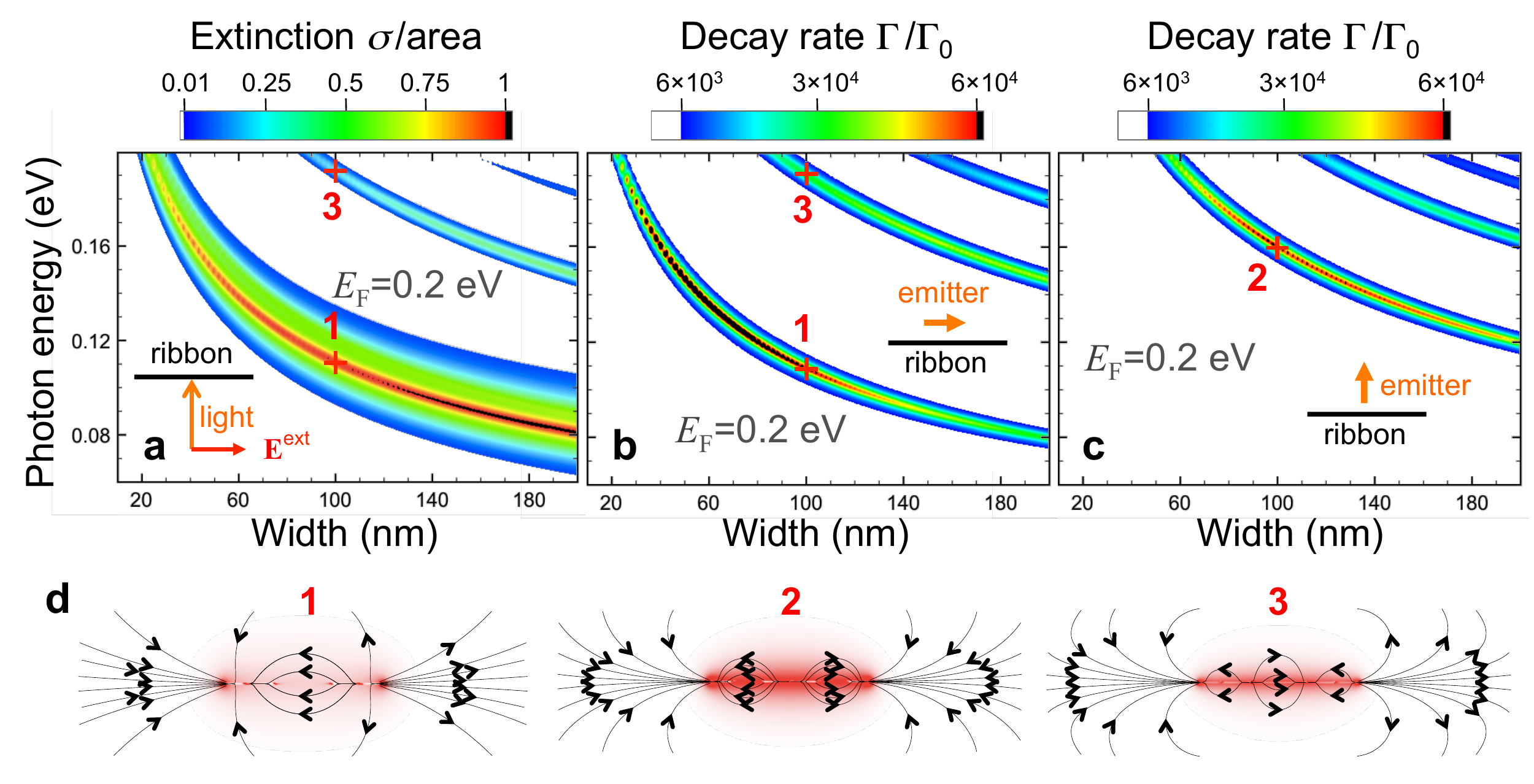}}
\caption{{\bf Resonant coupling to graphene ribbons.}  {\bf a,} Extinction cross-section of doped self-standing graphene ribbons as a function of ribbon width and photon energy for a Fermi energy $E_F=0.2\,$eV. The light is incident as shown in the inset. The cross section is normalized to the carbon sheet area. {\bf b,c,} Decay rate normalized to free space under the same conditions as in {\bf a} for line emitters situated 10\,nm above the center of the ribbon and polarized parallel (b) or perpendicular (c) to it. {\bf d,} Near electric-field intensity and field lines for the modes corresponding to labels 1-3 in {\bf a,c}.} \label{Fig3}
\end{figure}

\begin{figure}
\centerline{\includegraphics*[width=10cm]{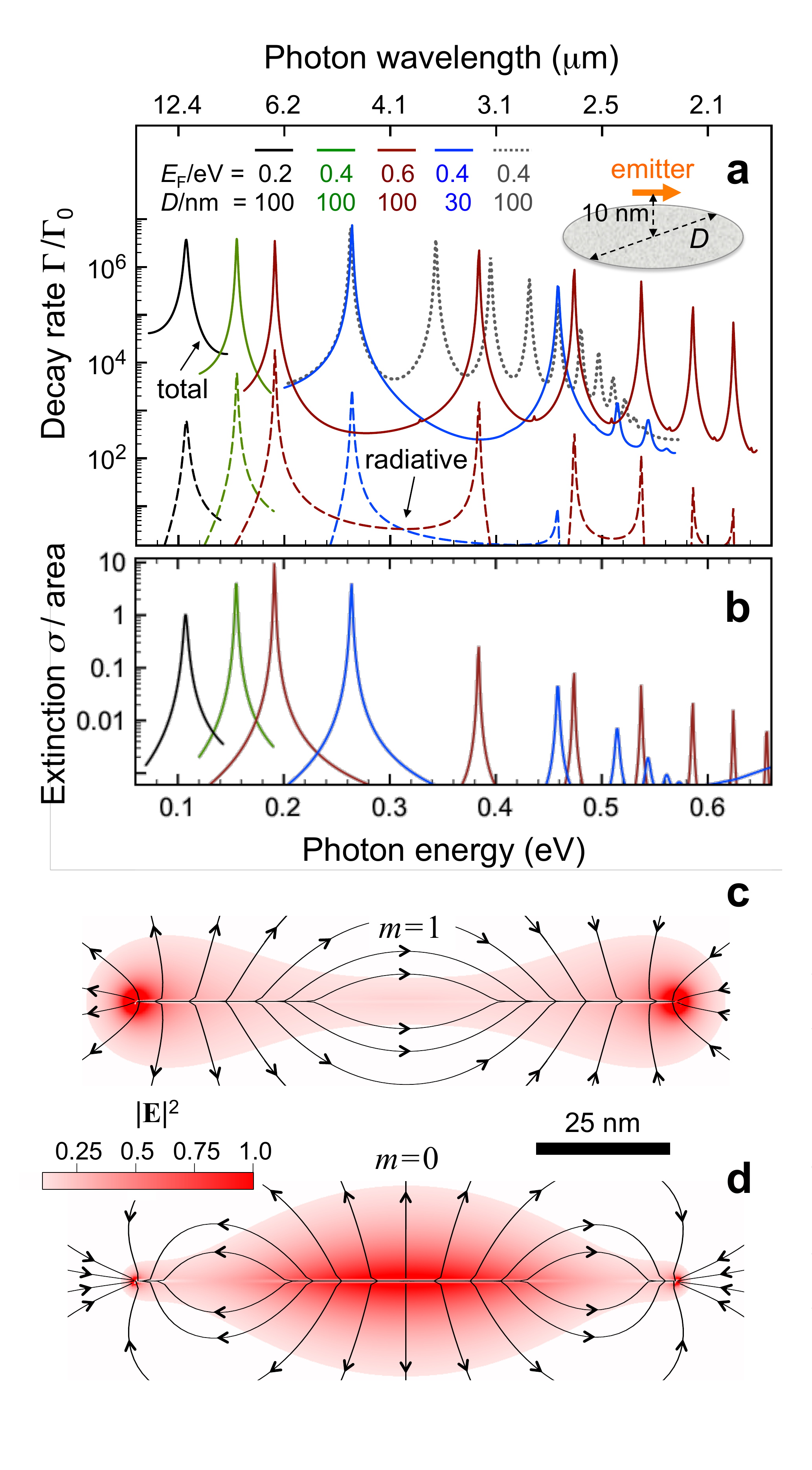}}
\caption{{\bf Plasmons in graphene nanodisks.} {\bf a,} Decay rate of an emitter situated 10\,nm above the center of a doped graphene disk for two different disk diameters $D$ and various values of the Fermi energy $E_F$, as shown in the legend. Total decay rates (solid curves) are compared to the contribution of radiation emission (dashed curves). The emitter is polarized parallel to the disk, so that it excites SPs of $m=1$ azimuthal symmetry, except in the dotted curve, corresponding to perpendicular orientation ($m=0$ symmetry) for $D=100\,$nm and $E_F=0.4\,$eV. {\bf b,} Normal-incidence extinction cross-section of the same disks as in {\bf a}. {\bf c,d,} Near-electric-field intensity of the lowest-energy modes in a $D=100\,$nm disk doped to $E_F=0.4\,$eV for $m=0$ ($\hbar\omega=0.16\,$eV) and $m=1$ ($\hbar\omega=0.26\,$eV) azimuthal symmetries. Electric field-lines are superimposed on the intensity plot.} \label{Fig4}
\end{figure}

\begin{figure}
\centerline{\includegraphics*[width=7cm]{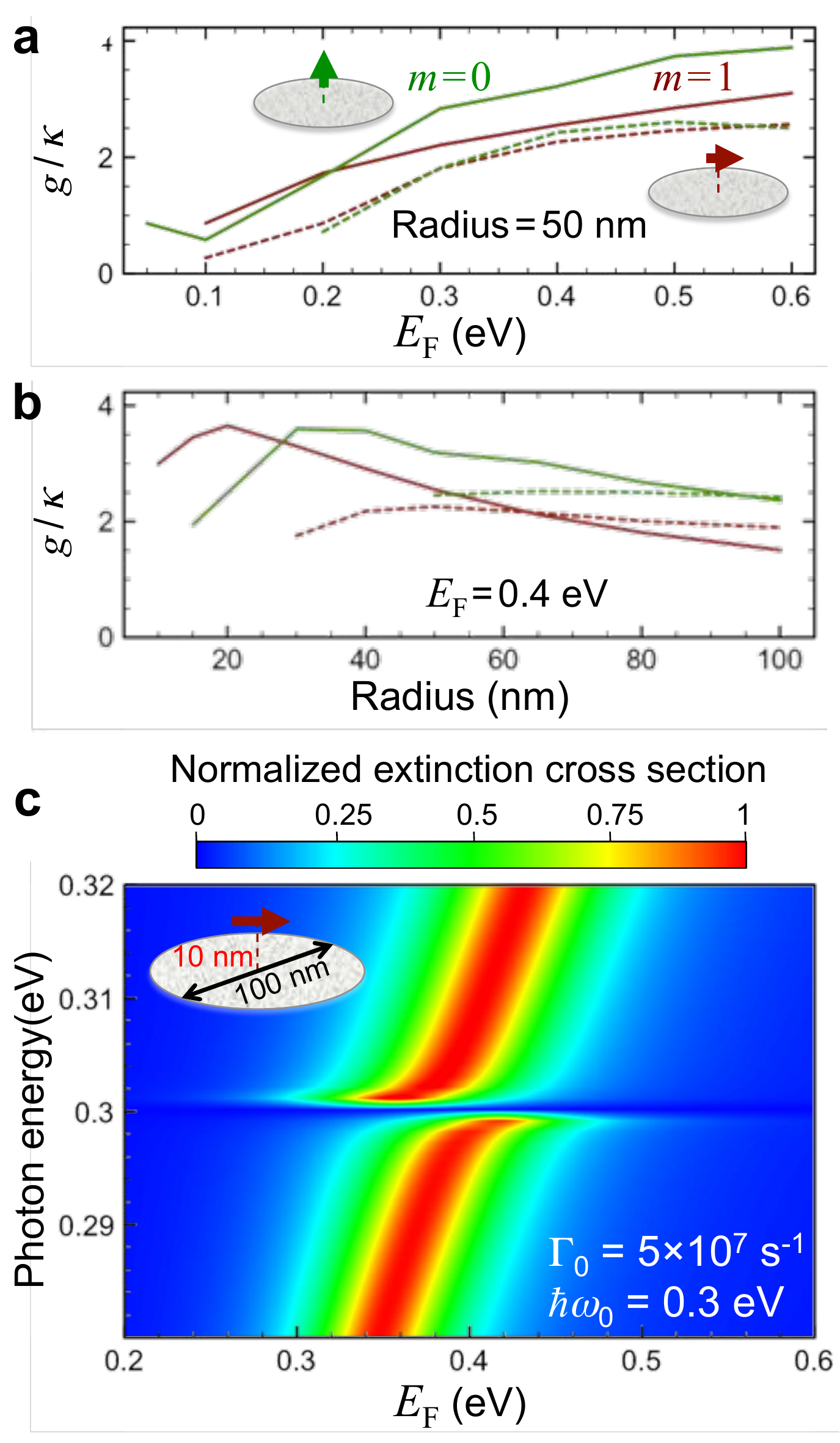}}
\caption{{\bf Strong-coupling and vacuum Rabi splitting in graphene nanodisks.} {\bf a,b,} Fermi-energy and disk-radius dependence of the strong-coupling parameter $g/\kappa$ for an emitter placed 10\,nm above the center of a doped graphene disk for the first-order (solid curves) and second-order (dashed curves) modes with either $m=0$ (green curves) or $m=1$ (red curves) azimuthal symmetries (the two lowest-order modes are shown for each symmetry). The natural decay rate of the emitter is $\Gamma_0=5\times10^7\,$s$^{-1}$ (see Eq.\ (\ref{Hint}) and below). {\bf c,} Fermi- and photon-energy dependence of the extinction cross-section of a combined emitter-nanodisk system under the same conditions as in {\bf a}. The emitter has a resonance at $\hbar\omega_0=0.3\,$eV and is oriented parallel to the disk. The cross section is normalized to the maximum resonant extinction of the isolated disk.} \label{Fig5}
\end{figure}

\clearpage

\section{Appendix for\\"Graphene Plasmonics: A Platform for Strong Light-Matter Interaction"}
\subsection{Simulations and relaxation time}

\indent

{\bf Electromagnetic simulations.} We describe ribbons by expanding the induced current in Fourier series, assuming that the parallel external field is directed across the ribbon, using a supercell with sufficiently spaced carbon sheets (see Appendix). Good convergence is obtained with 400 Fourier components and supercell spacings of 4 ribbon widths. This method produces excellent agreement with an alternative approach fully relying on numerical electromagnetic simulations consisting in modeling the graphene as a thin film of dielectric function $1+4\pi i\sigma/\omega t$ and thickness $t=0.5\,$nm, with the edges rounded by hemi-circular profiles, for which we find converged results using the boundary element method (BEM), as shown in the Appendix \cite{paper070}. Nanodisks are also simulated with the BEM. The conductivity is taken from the $k_\parallel\rightarrow0$ limit of the RPA\cite{FV07} in all cases. The decay rates are obtained from the self-induced field of a dipole using Eq.(3).

{\bf Relaxation time.} This is an important parameter because the actual value of $\tau$ affects the plasmon propagation distance, which is nearly proportional to $1/\tau$, although the decay rates computed here are rather insensitive to $\tau$ (because they are $k_\parallel$-integrated quantities), except near the intraband onset (see Appendix). We estimate $\tau$ from the measured, impurity-limited DC mobility\cite{NGM04,NGM05} $\mu\approx10,000\,$cm$^2/$Vs, which yields $\tau=\mu E_F/ev_F^2\approx10^{-13}\,$s for $E_F=0.1\,$eV, to be compared with $\sim10^{-14}\,$s in gold. We note that this is a very conservative value compared to recent observations in high-quality suspended graphene\cite{BSJ08} ($\mu>100,000$) and graphene on boron nitride\cite{DYM10} ($\mu=60,000$). Optical phonons are known to contribute to $\tau$ above the phonon frequency $\sim0.2\,$eV. Careful analysis \cite{JBS09} reveals that their effect can be incorporated through an effective $\tau\sim0.5\times10^{-13}\,$s, which produces a reduction in the peak decay rates comparable to the increase in $1/\tau$ (e.g., by a factor of 5-10 in the spectra of Fig.\ 4, see Appendix).

\noindent

\subsection{Graphene conductivity in the random-phase approximation (RPA)}

The nonlocal conductivity is related to the susceptibility through
\begin{equation}
\sigma(k_\parallel,\omega)=-i\omega\chi_\tau(k_\parallel,\omega). \nonumber
\end{equation}
We introduce a finite relaxation time $\tau$ using the prescription given by Mermin \cite{M1970,JBS09}, which preserves the number of charge carriers:
\begin{equation}
\chi_\tau(k_\parallel,\omega)=\frac{(1+i/\omega\tau)\chi(k_\parallel,\omega+i/\tau)}{1+(i/\omega\tau)\chi(k_\parallel,\omega+i/\tau)/\chi(k_\parallel,0)}, \nonumber
\end{equation}
where
\begin{equation}
\chi(k_\parallel,\omega)=\frac{e^2}{2\pi^2\hbar k_\parallel^2}
\int d^2{\bf k}'_\parallel\sum_{s,s'=\pm}\left[1+ss'\frac{{\bf k}'_\parallel\cdot({\bf k}_\parallel+{\bf k}'_\parallel)}{k^\prime_\parallel|{\bf k}_\parallel+{\bf k}'_\parallel|}\right]\frac{\theta_F(s'v_F|{\bf k}_\parallel+{\bf k}'_\parallel|)-\theta_F(sv_Fk'_\parallel)}{\omega+v_F\left(sk'_\parallel-s'|{\bf k}_\parallel+{\bf k}'_\parallel|\right)+i0^+}\nonumber
\end{equation}
is the linear RPA response function \cite{WSS06,HD07} and $\theta_F(E)$ is the Fermi-Dirac distribution.

The RPA response admits an analytical expression at zero temperature \cite{WSS06} (i.e., for $\theta_F(E)=\theta(E_F-E)$):
\begin{equation}
\chi(k_\parallel,\omega)=\frac{e^2}{4\pi\hbar}\left[\frac{8k_F}{v_Fk_\parallel^2}
\;\;+\;\;\frac{G(-\Delta_-)\;\theta\left[-{\rm Re}\{\Delta_-\}-1\right]
+\left[G(\Delta_-)+i\pi\right]\;\theta\left[{\rm Re}\{\Delta_-\}+1\right]-G(\Delta_+)}{\sqrt{\omega^2-v_F^2k_\parallel^2}}\right],\nonumber
\end{equation}
where
\begin{equation}
G(z)=z\sqrt{z^2-1}-\log\left(z+\sqrt{z^2-1}\right)\nonumber
\end{equation}
and $\Delta_\pm=(\omega/v_F\pm2k_F)/k_\parallel$. Here, the square roots are chosen to yield positive real parts, while the imaginary part of the logarithm is taken in the $(-\pi,\pi]$ range. Additionally, we have 
\begin{equation}
\chi(k_\parallel,0)=\frac{e^2}{2\pi\hbar v_Fk_\parallel}\left\{\frac{4k_F}{k_\parallel}
-\;\theta(1-x)\left[x\sqrt{1-x^2}-\cos^{-1}x\right]\right\},\nonumber
\end{equation}
where $x=2k_F/k_\parallel$. We use these formulas to compute the nonlocal RPA in this document.

To a good approximation (see below) the conductivity can be evaluated within the local RPA (i.e., the $k_\parallel\rightarrow0$ limit), which leads to an analytical solution including the dependence on $T$ \cite{FV07}:
\begin{align}
\sigma(\omega)=&\frac{2e^2T}{\pi\hbar}\frac{i}{\omega+i\tau^{-1}}\log\left[2\cosh(E_F/2k_BT)\right]
\label{localRPA}\\
&+\frac{e^2}{4\hbar}\left[H(\omega/2)+\frac{4i\omega}{\pi}
\int_0^\infty d\varepsilon\;\frac{H(\varepsilon)-H(\omega/2)}{\omega^2-4\varepsilon^2}
\right], \nonumber
\end{align}
where
\begin{equation}
H(\varepsilon)=\frac{\sinh(\hbar\varepsilon/k_BT)}{\cosh(E_F/k_BT)+\cosh(\hbar\varepsilon/k_BT)}. \nonumber
\end{equation}
The first term in Eq.\ (\ref{localRPA}) corresponds to intra-band transitions, in which the relaxation time has been introduced to make it converge to the Drude model at $T=0$. We show below that nonlocal effects produce qualitatively similar results as a finite relaxation time $\tau$ and temperature $T$. Because the actual value of $\tau$ depends on the quality of the synthetized graphene, it can actually be regarded as an effective parameter. Actually, the decay rate of an emitter in the vicinity of homogeneous graphene is rather insensitive to the inclusion of nonlocal effects and the actual value of $\tau$ and $T$ within the wide spectral region for which the plasmons are well defined (see below). Therefore, we use Eq.\ (\ref{localRPA}) for the conductivity in all calculations presented here and in the main paper (unless it is stated otherwise) because it gives a reasonable description and it is local, so that we assume that it can be also used for patterned graphene.

When $T=0$, Eq.\ (\ref{localRPA}) reduces to
\begin{equation}
\sigma(\omega)=\frac{e^2E_F}{\pi\hbar^2}\frac{i}{\omega+i\tau^{-1}}
+\frac{e^2}{4\hbar}\left[\theta(\hbar\omega-2E_F)+\frac{i}{\pi}
\log\left|\frac{\hbar\omega-2E_F}{\hbar\omega+2E_F}\right|
\right],\label{localRPAT0}
\end{equation}
which shows a sudden increase in losses (step function affecting the real part of $\sigma$) at the onset of vertical intra-band transitions, $\hbar\omega=2E_F$. Full inclusion of finite temperature and damping leads to a smoother onset, but Eq.\ (\ref{localRPAT0}) contains the main features of the graphene conductivity.

\begin{figure}
\includegraphics[width=140mm,angle=0]{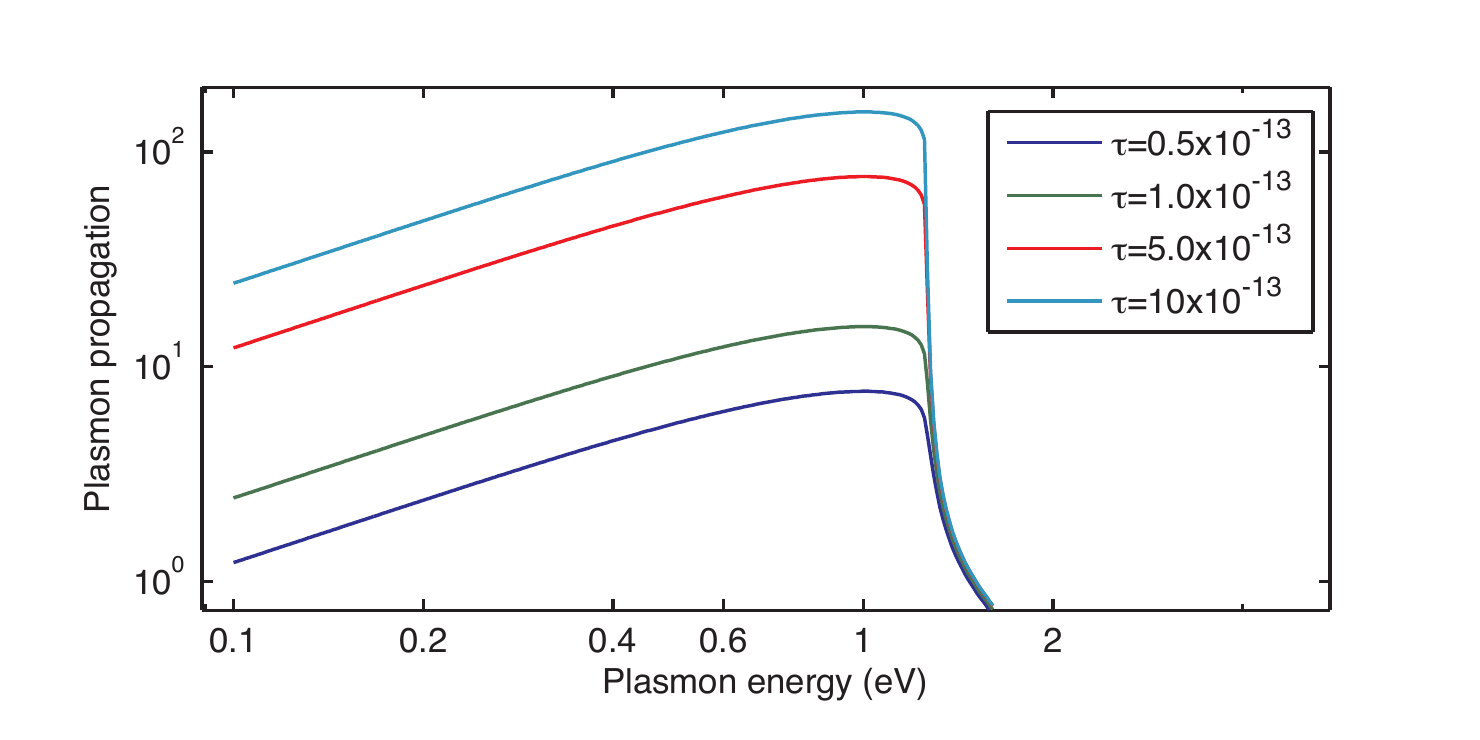}
\caption{Spectral dependence of the in-plane plasmon propagation distance (in units of the plasmon wavelength) obtained from the RPA for various relaxation times in homogeneous graphene supported on an $\epsilon=2$ material and doped to $E_F=1\,$eV.} \label{FigSI1}
\end{figure}

\subsection{Fresnel coefficients and plasmon dispersion in homogeneous graphene}
\label{SecFresnel}

The response of homogeneous graphene is expressed in terms of its Fresnel reflection coefficients \cite{NH06}, which can be obtained by applying the customary boundary conditions ($\Delta {\bf E}_\parallel=\Delta {\bf H}_\perp=0$, $\Delta(\epsilon{\bf E}_\perp)=4\pi\sigma\nabla\cdot{\bf E}_\parallel/i\omega$, and $\hat{\bf n}\times\Delta{\bf E}_\perp=(4\pi\sigma/c){\bf E}_\parallel$) for the fields of incoming $p$- and $s$-polarized plane waves as
\begin{align}
r_p&=\frac{\epsilon k_\perp-k'_\perp+4\pi\sigma k_\perp k'_\perp/\omega}{\epsilon k_\perp+k'_\perp+4\pi\sigma k_\perp k'_\perp/\omega}, \label{rp}\\
r_s&=\frac{k_\perp-k'_\perp+4\pi\sigma k_0/c}{k_\perp+k'_\perp+4\pi\sigma k_0/c}, \nonumber
\end{align}
where $\epsilon$ is the permittivity of the substrate on which the graphene is deposited, $k_0=\omega/c$ is the free-space light wave vector, $k_\perp=\sqrt{k_0^2-k_\parallel^2}$ and $k'_\perp=\sqrt{\epsilon k_0^2-k_\parallel^2}$ are the perpendicular wave vectors outside and inside the substrate, respectively, and $k_\parallel$ is the parallel wave vector.

The dispersion relation of $p$-polarized surface plasmons (SPs) is determined by the pole of $r_p$, which yields the equation \[\epsilon/\sqrt{\epsilon k_0^2-k_{\rm sp}^2}+1/\sqrt{k_0^2-k_{\rm sp}^2}=-4\pi\sigma/\omega\] for the plasmon wave vector $k_{\rm sp}$. Here, we can use the electrostatic limit of this expression, \[k_{\rm sp}\approx i(\epsilon+1)\omega/4\pi\sigma,\]
under the common condition $k_0\ll|k_{\rm sp}|$ (actually, this condition is fulfilled for $\hbar\omega\gg\alpha E_F$, where $\alpha\approx1/137$ is the fine-structure constant).

We plot the plasmon dispersion relation in Fig.\ 1 of the main paper, and we supplement it here by showing the $1/e$-amplitude-decay propagation length computed from $1/{\rm Im}\{k_{\rm sp}\}$ as a function of plasmon energy for various values of $\tau$. For the relaxation times considered in this work, the susceptibility has an almost linear dependence on the damping rate $1/\tau$ that translates into a linear variation of the plasmon propagation length with this parameter, as shown in Fig.\ \ref{FigSI1} for $E_F=1\,$eV.

\begin{figure}
\includegraphics[width=130mm,angle=0]{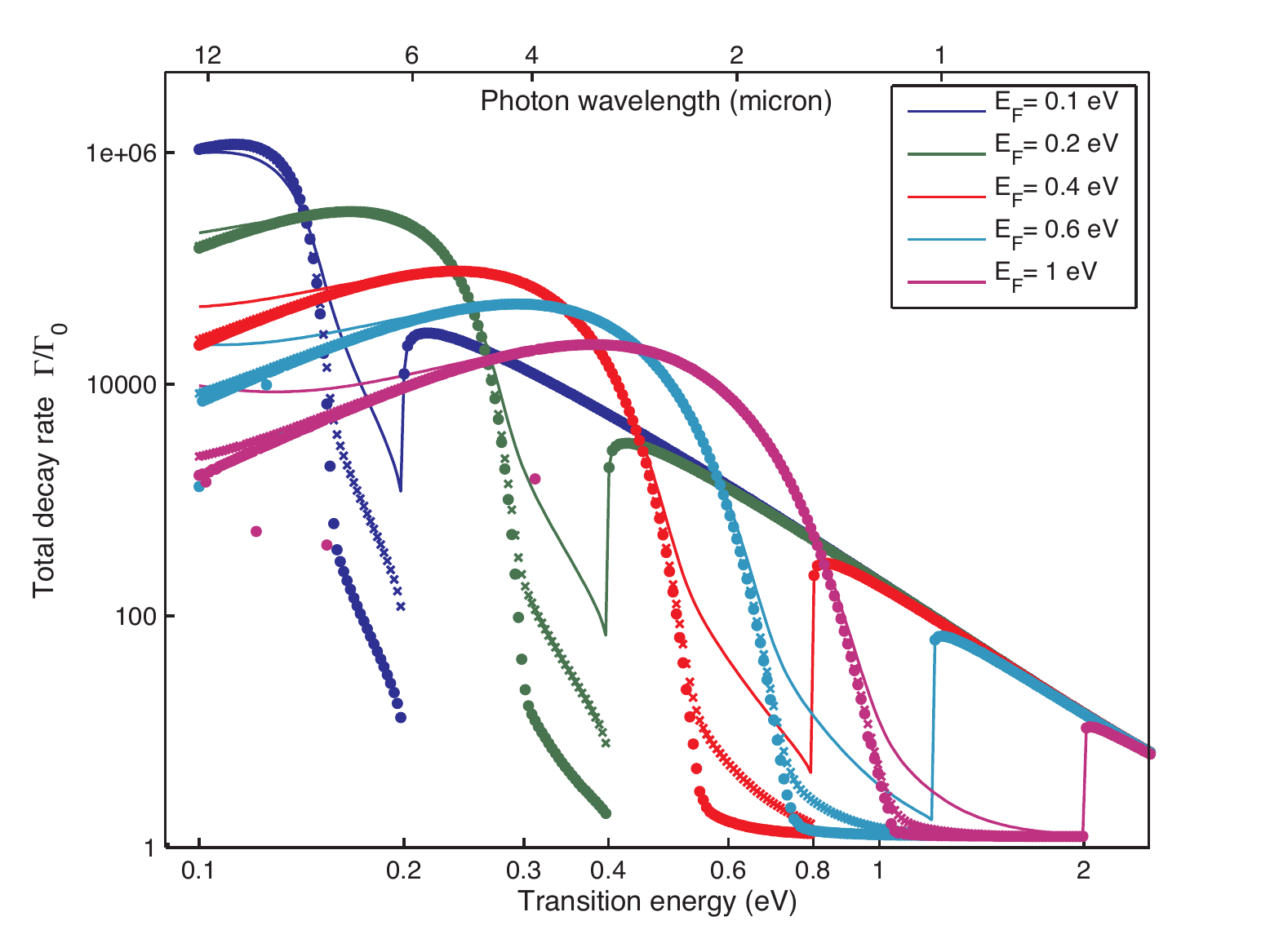}
\caption{Decay rate calculated from the RPA at $T=0$ for $\tau=50\,$fs (solid curves), $\tau=500\,$fs (crosses), and  $\tau=5,000\,$fs (dots). The emitter is placed 10\,nm away from a homogeneous graphene sheet deposited on the surface of an $\epsilon=2$ material. The rate $\Gamma$ is normalized to the free-space rate $\Gamma_0$.} \label{FigSI2}
\end{figure}
\begin{figure}
\includegraphics[width=130mm,angle=0]{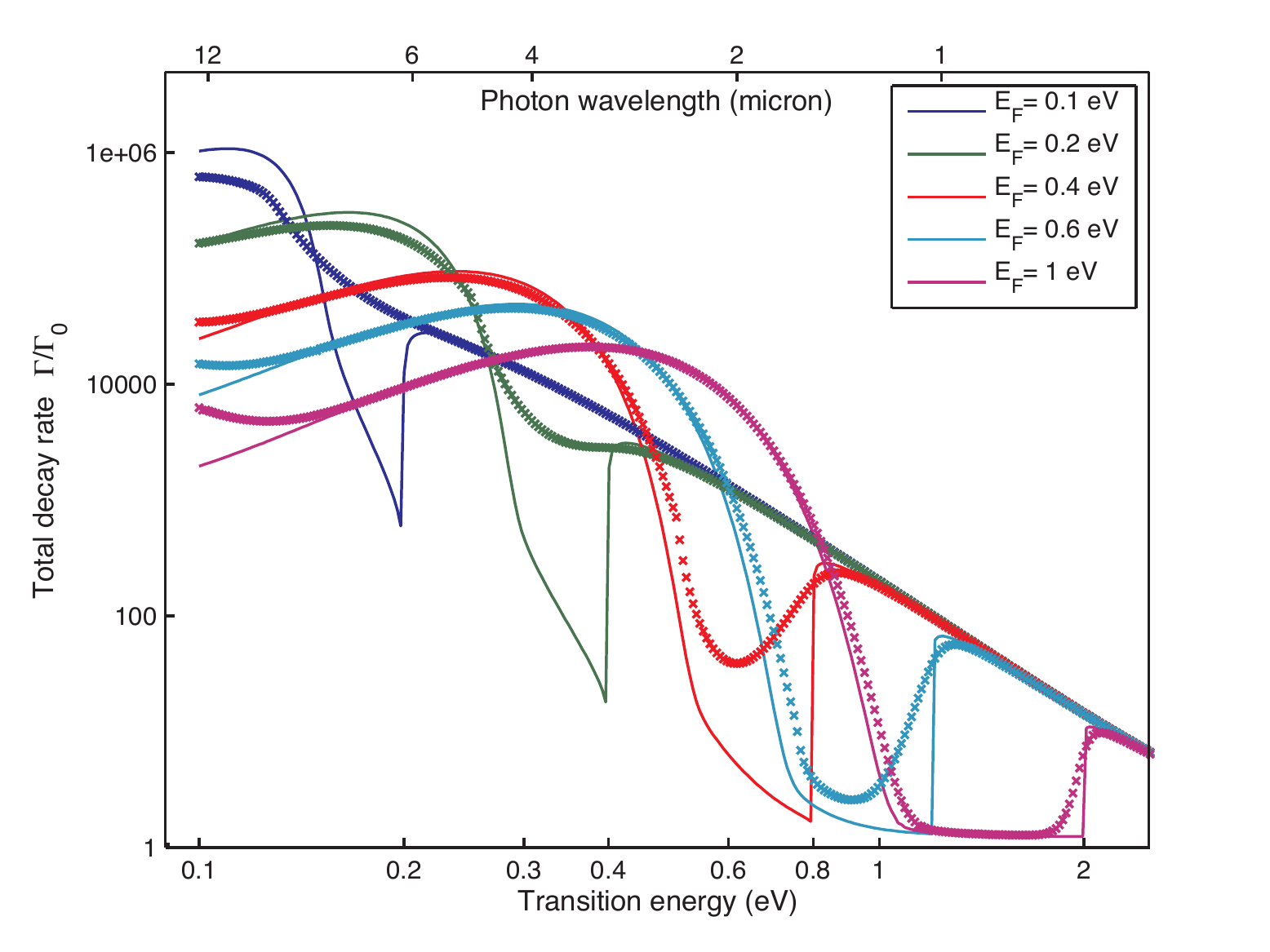}
\caption{Decay rate obtained from the RPA (symbols) and the local RPA (curves) under the same conditions as in Fig.\ \ref{FigSI2} with $T=0$, $\tau=\mu E_F/ev_F^2$, and a mobility $\mu=10,000\,$cm$^2/$Vs (e.g., $\tau\approx10^{-13}\,$s for $E_F=0.1\,$eV).} \label{FigSI3}
\end{figure}
\begin{figure}
\includegraphics[width=150mm,angle=0]{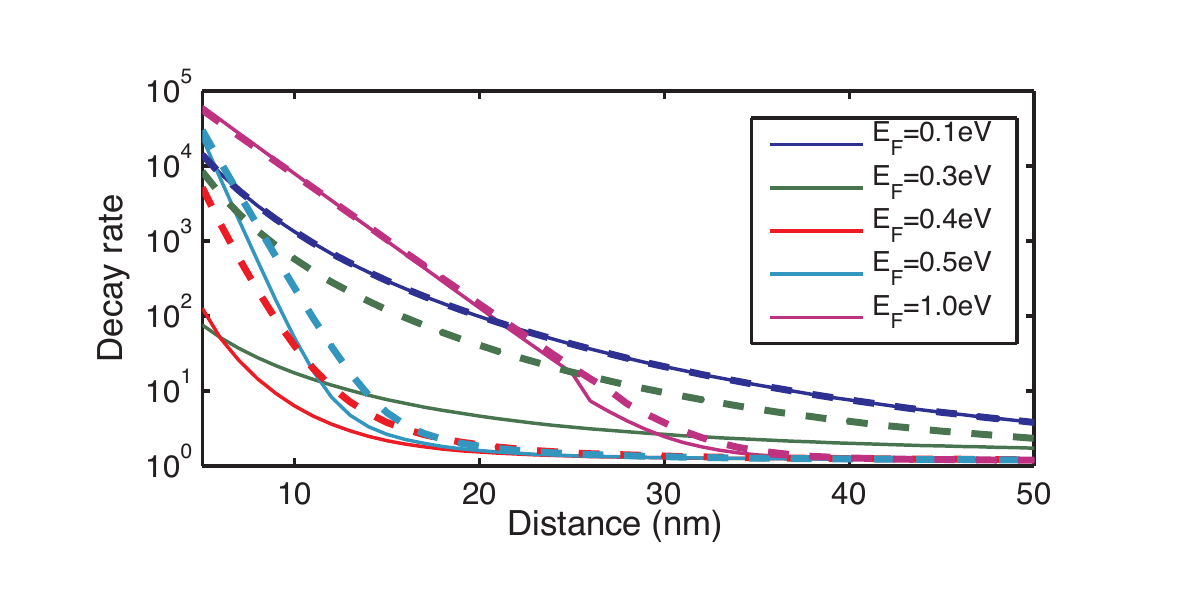}
\caption{Distance dependence of the decay rate under the conditions of Fig.\ \ref{FigSI3} for a photon energy of $0.5\,$eV. Solid curves: local RPA. Dashed curves: RPA.} \label{FigSI4}
\end{figure}
\begin{figure}
\includegraphics[width=130mm,angle=0]{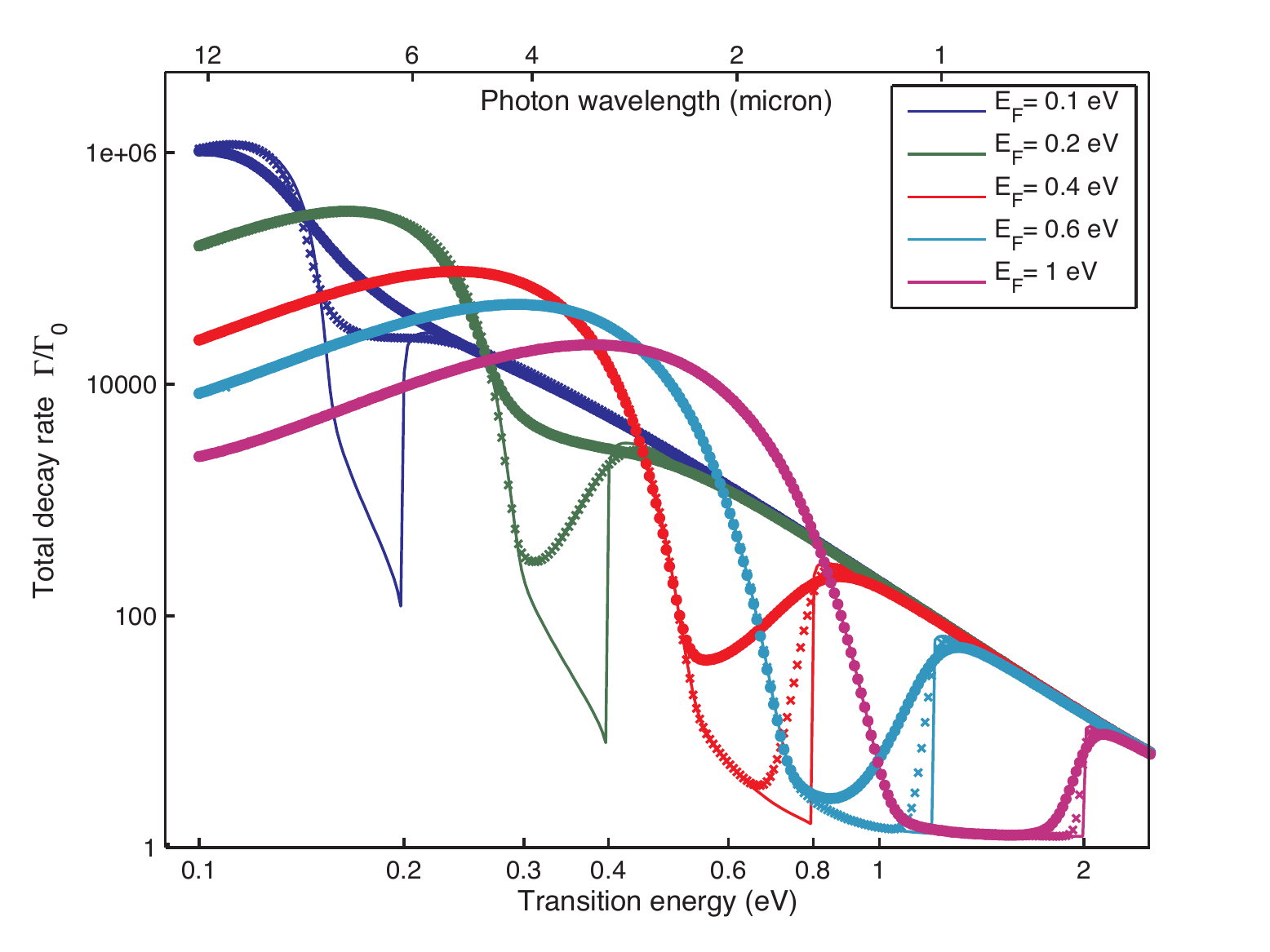}
\caption{Decay rate calculated from the local RPA at $T=0$ (solid curves), $T=100$ (crosses), and $T=300$ (dots) for $\tau=500\,$fs under the same conditions as in Fig.\ \ref{FigSI2}.} \label{FigSI5}
\end{figure}

\subsection{Decay rate and its dependence on conductivity model, temperature, and relaxation time}

The decay rate $\Gamma$ can be related to the electric field induced by a dipole $\db$ on itself $\Eb^{\rm ind}$ as \cite{NH06}
\begin{equation}
\Gamma=\Gamma_0+\frac{2}{\hbar}{\rm Im}\{\db^*\cdot\Eb^{\rm ind}\}, \label{GamGam}
\end{equation}
where $\Gamma_0=4k_0^3|\db|^2/3\hbar$ is the free-space decay rate. When the emitter is above a substrate covered with a homogeneous graphene layer, the induced or reflected field can be in turn related to the Fresnel coefficients of graphene to yield \cite{NH06}
\begin{equation}
\Gamma=\Gamma_0+\frac{1}{\hbar}\int_0^\infty k_\parallel\,dk_\parallel\,{\rm Re}\left\{\left[|{\bf d}_\parallel|^2(k_0^2r_s-k_\perp^2r_p)+2|{\bf d}_\perp|^2 k_\parallel^2r_p\right]\,\frac{e^{2ik_\perp z}}{k_\perp}\right\}, \label{Gamplanar}
\end{equation}
where $z$ is the emitter-graphene separation, ${\bf d}_\parallel$ and ${\bf d}_\perp$ are the components of the transition dipole parallel and perpendicular to the carbon plane, and the integral is extended over parallel wave vectors $k_\parallel$. In this work, we use Eq.\ (\ref{Gamplanar}) to compute the decay rate in homogeneous graphene, and Eq.\ (\ref{GamGam}) for nanoribbons and nanodisks, with $\Eb^{\rm ind}$ calculated as explained in Sec.\ \ref{sec4} and \ref{sec5}.

The decay rate $\Gamma$ is a $k_\parallel$-integrated quantity, and therefore, we expect a mild dependence on the relaxation time $\tau$, except in the neighborhood of the onset of vertical inter-band transitions, where $\Gamma$ takes small values that are incremented by the smearing of the electron-hole pair (e-h) continuum due to relaxation. This is the conclusion that can be extracted from Fig.\ \ref{FigSI2}, in which we plot the spectral dependence of $\Gamma$ for various values of the Fermi energy and we consider a wide range of relaxation parameters. The rate is nearly independent of $\tau$ over the region of existence of surface plasmons and also above the noted onset, where it convergences to the undoped graphene level.

A similar conclusion can be drawn when comparing the RPA with the local RPA (Fig.\ \ref{FigSI3}). They produce nearly the same results, except in the spectral region extending from the plasmon cutoff to the vertical inter-band transition threshold. Clearly, the agreement in the plasmon region improves for higher $E_F$, presumably as a result of the momentum cutoff for finite separation between the emitter and the graphene, which effectively reduces the effect of non-vertical e-h transitions. The two models also differ in the low-energy region. The agreement in the plasmonic region is also observed in the distance dependence of $\Gamma$ (Fig.\ \ref{FigSI4}), although severe discrepancies are observed for $E_F$ slightly below the photon energy.

We explore the variation with temperature in Fig.\ \ref{FigSI5}. The effect of a finite temperature is similar to that of finite relaxation, essentially consisting in smearing the dip in the decay rate below the $2E_F$ onset.

\subsection{Fourier expansion method for nanoribbons}
\label{sec4}

We consider a nanoribbon contained in the $z=0$ plane and having translational invariance along $y$. Furthermore, we assume the external field to be independent of $y$. This is the case for illumination with a plane wave normal to the graphene and polarized along $x$ (actually, this is the geometry for which we calculate the cross section in this work), and also for emission from a line dipole polarized along either $x$ or $z$ and consisting of a continuous distribution of identical point dipoles distributed along a line parallel to $y$ (we obtain decay rates for this configuration).

Under these conditions, the component of the electric field parallel to the graphene is along $x$, and thus, the induced current $\eta(x)$ is also along $x$. The field produced by each surface element $dxdy$ is the same as that of a dipole $(i\eta/\omega)\,dxdy$. Summing all of these dipole contributions, and including the effect of a substrate through its Fresnel coefficients \cite{NH06}, we find the self-consistent relation
\begin{equation}
\eta(x)/\sigma=E_x^{\rm ext}(x)-\frac{1}{\omega}\int dq\;k_\perp (1-r_p) \int dx' e^{iq(x'-x)}\eta(x'), \label{etaself}
\end{equation}
where the $x$ integral is extended over the graphene width, $k_\perp=\sqrt{k_0^2-q^2}$, $k_0$ is the free-space light wave vector, and $r_p$ is the Fresnel coefficient of the substrate for $p$ polarization. More precisely, $r_p=(\epsilon k_\perp-k'_\perp)/(\epsilon k_\perp+k'_\perp)$, where $k'_\perp=\sqrt{\epsilon k_0^2-q^2}$. Here, $E_x^{\rm ext}$ is the external electric field along $x$, which already includes the reflection by the homogeneous dielectric substrate of permittivity $\epsilon$.

We solve Eq.\ (\ref{etaself}) by considering a supercell with the graphene occupying the $z=0$ and $0<x<b$ region, and by periodically repeating this unit cell with period $a$ along $x$. Then, we expand the conductivity, the external field, and the surface current in Fourier series. For example, the conductivity becomes \[\sigma(x)=\sum_n \sigma_n e^{ig_nx}\] (it is zero outside the graphene and given by Eq.\ (\ref{localRPA}) in the graphene), where $g_n=2\pi n/a$, and the coefficients $\sigma_n$ are easily obtained from the expansion of the step function representing the ribbon. This allows us to project Eq.\ (\ref{etaself}) into Fourier components as
\begin{equation}
\eta_n=\frac{1}{a}\int_0^bdx\,\sigma(x)\,E_x^{\rm ext}(x)\,e^{-ig_nx}-\frac{2\pi}{\omega}\sum_{n'}\;k^{n'}_\perp (1-r^{n'}_p) \sigma_{n-n'}\eta_{n'}, \label{four}
\end{equation}
where $k^{n'}_\perp$ and $r^{n'}_p$ are the same as $k_\perp$ and $r_p$ for $q=g_{n'}$. Finally, we solve Eq.\ (\ref{four}) by using standard linear algebra with a finite number of waves $M$.

The scattered near-field is given in terms of the $\eta_n$ coefficients as
\begin{equation}
{\bf E}^{\rm scat}=-\frac{2\pi}{\omega}\,\sum_n\,\eta_n\,e^{ig_nx}\times
\begin{cases}
e^{ik^n_\perp z} (1-r^n_p)\;(k^n_\perp,0,-g_n), & z>0, \\
& \\
e^{-i{k'}^n_\perp z} (1+r^n_p)\;({k'}^n_\perp,0,g_n)/\epsilon, & z<0,
\end{cases}\nonumber
\end{equation}
where the substrate is taken to occupy the $z<0$ region.

We obtain convergence for isolated ribbons by taking $a$ equal to 2-4 times $b$. Then, we derive the single-ribbon far-field from the induced current of the ribbon in the first unit cell:
\begin{equation}
{\bf E}^{\rm scat}=\frac{e^{ik'R}}{\sqrt{k'R}}\,f\,\hat\varepsilon_p,\nonumber
\end{equation}
where $\hat\varepsilon_p$ is the unit vector for scattered $p$-polarized light,
\begin{equation}
f=\frac{\sqrt{2\pi} k_0}{c}e^{i\pi/4}(1\pm r_p)\;\cos\theta\;\sum_n\eta_nI_n  \label{ff}
\end{equation}
is the field amplitude, $\theta$ is the angle relative to the outwards surface normal, $k'=k_0$ ($k'=k_0\sqrt{\epsilon}$) above (inside) the substrate, the upper (lower) sign in (\ref{ff}) applies outside (inside) the substrate, and
\begin{equation}
I_n=\frac{e^{i(g_n-q)b}-1}{g_n-q}. \nonumber
\end{equation}
Here, the reflection coefficient $r_p$ and the parallel wave vector $q=k_0\sin\theta$ are determined by the outgoing angle $\theta$.

Applying the optical theorem to the transmitted and reflected light upon plane wave illumination and using the above expressions for the far field, we find the extinction cross section
\begin{equation}
\sigma^{\rm ext}=\frac{8\pi}{\omega}\frac{k_\perp k'_\perp}{\epsilon k_\perp+k'_\perp}\;{\rm Im}\left\{-\sum_n\eta_nI_n\right\}. \nonumber
\end{equation}

Finally, the decay rate per unit length along $y$ for a line dipole of per-unit-length strength ${\bf d}$ ($\perp y$) situated above the substrate is calculated from
\begin{equation}
\Gamma=\frac{\pi k_0^2d^2}{\hbar}+\frac{2}{\hbar}{\rm Im}\{{\bf E}^{\rm ind}\cdot{\bf d}\}, \nonumber
\end{equation}
where the induced field is evaluated at the position of the dipole.

\subsection{Convergence of Fourier expansion and boundary element method}
\label{sec5}

\begin{figure}
\includegraphics[width=120mm,angle=0]{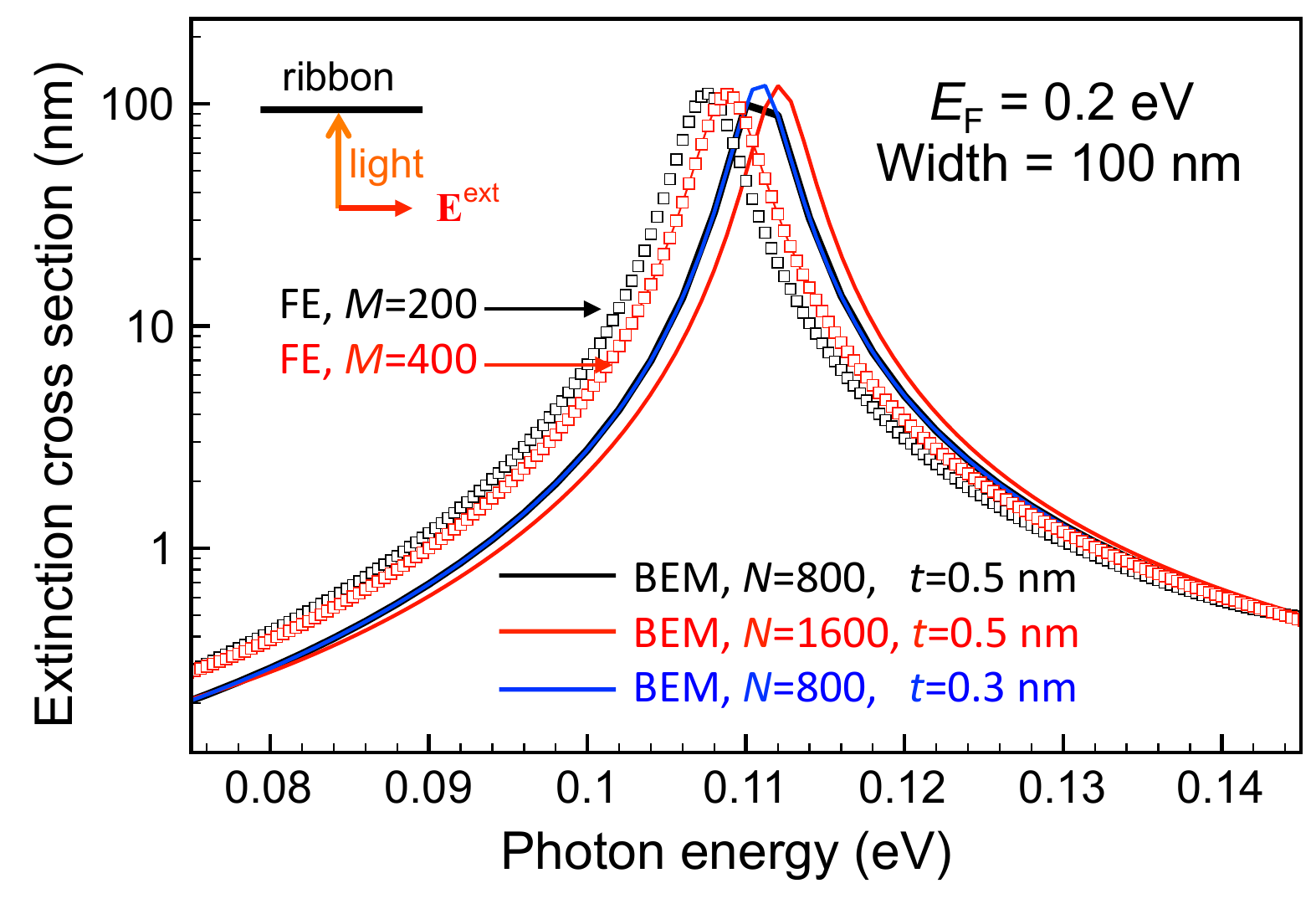}
\caption{Convergence of the Fourier expansion (FE) and the boundary element method (BEM) for a graphene nanoribbon of width 100\,nm and $E_F=0.2\,$eV. We represent the extinction cross section calculated with various values of the convergence parameters for normally-incident light. The number of Fourier components $M$ and boundary parametrization points $N$ are given in the text labels.} \label{FigSI6}
\end{figure}

The Fourier expansion method converges when the number of Fourier coefficients is increased, as we show in Fig.\ \ref{FigSI6} (symbols). This method produces results in excellent agreement with an alternative approach fully relying on numerical simulations, consisting in modeling the graphene as a thin film of dielectric function $1+4\pi i\sigma/\omega t$ and thickness $t$, with the edges rounded by hemi-circular profiles, for which we find converged electromagnetic results using the boundary element method (BEM) \cite{paper070}, as shown in Fig.\ \ref{FigSI6} (solid curves).

The agreement between the semi-analytical Fourier expansion and the BEM confirms the validity of the latter to describe graphene as a thin effective layer of dielectric. Actually, we use this method to simulate graphene disks, because an analytical expansion for them becomes too involved and does not add much insight into the problem.

\section{Supported vs self-standing nanoribbons}

\begin{figure}
\includegraphics[width=180mm,angle=0]{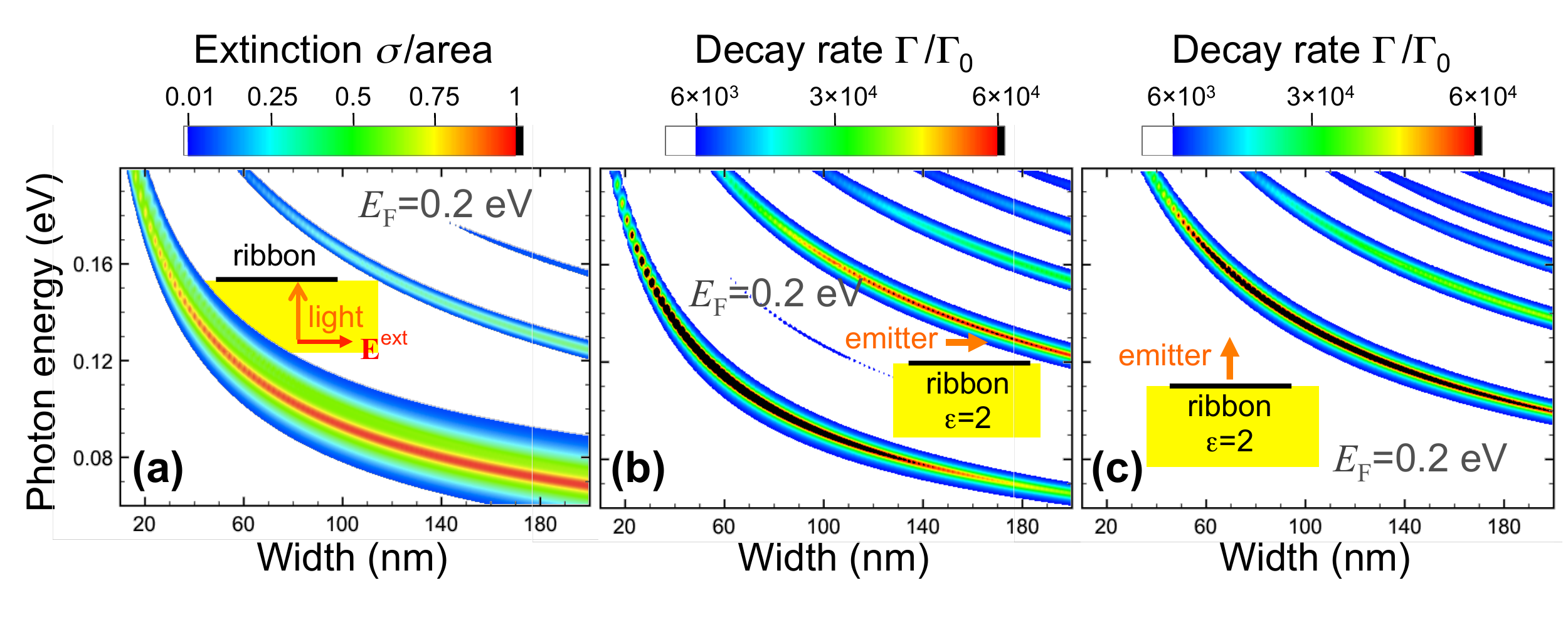}
\caption{{\bf (a)} Extinction cross section of doped graphene ribbons deposited on an $\epsilon=2$ material as a function of ribbon width and photon energy for a Fermi energy $E_F=0.2\,$eV. The light is incident as shown in the inset. The cross section is normalized to the carbon sheet area. {\bf (b,c)} Decay rate normalized to free space under the same conditions as in (a) for a line emitter situated 10\,nm above the center of the ribbon and polarized either parallel (b) or perpendicular (c) to it.} \label{FigSI7}
\end{figure}

We show in Fig.\ \ref{FigSI7} calculations similar to those of Fig.\ 3 of the main paper, but for graphene ribbons supported on the surface of an $\epsilon=2$ material rather than self-standing. The results are qualitatively the same for supported and for self-standing graphene. The extinction cross section and the decay rates have similar magnitude in both cases. The only difference that is worth noticing is the redshift in the plasmon energy in the supported graphene, which is consistent with the scaling of $\omega\propto1/\sqrt{\epsilon+1}$ predicted by the Drude model.

\subsection{Distance dependence of the decay rate near a nanodisk}

\begin{figure}
\includegraphics[width=180mm,angle=0]{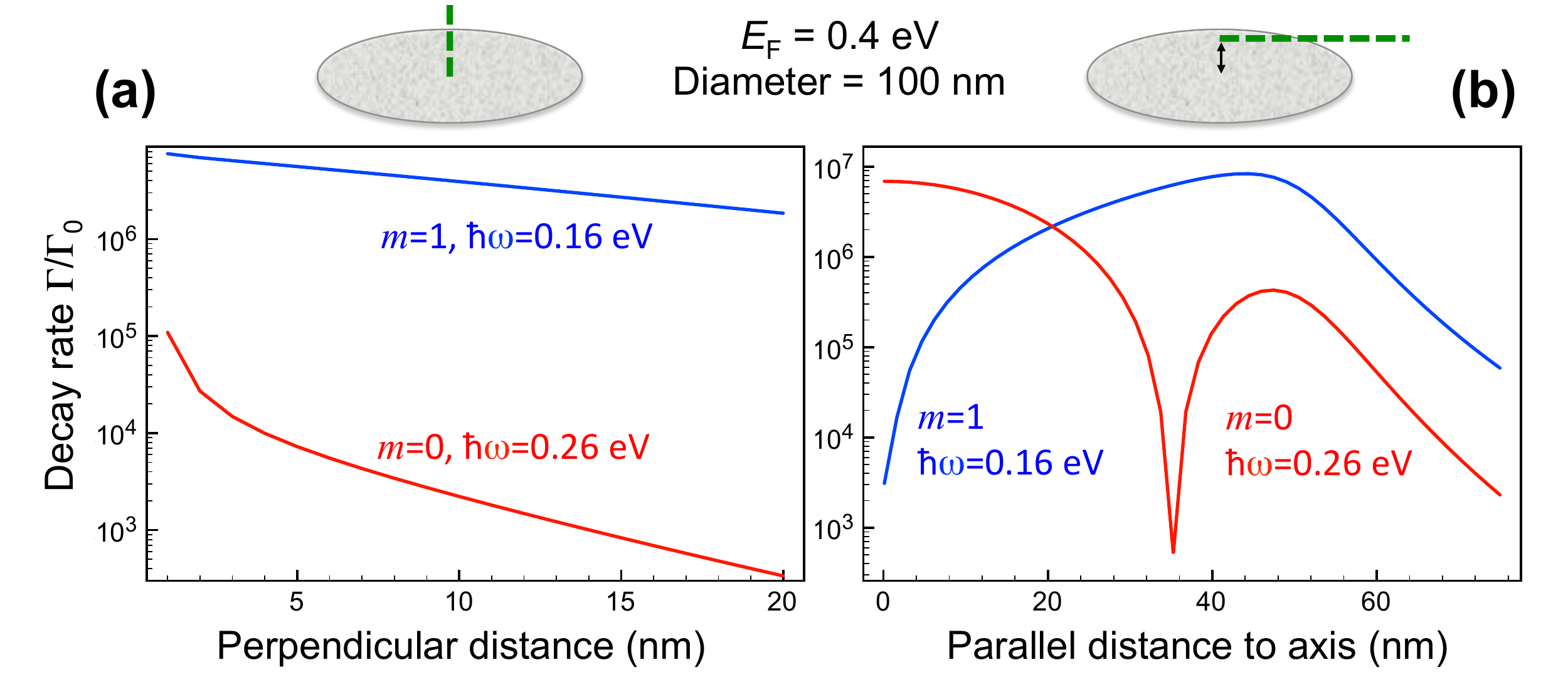}
\caption{Position dependence of the decay rate $\Gamma$ near a self-standing graphene disk of diameter 100\,nm and $E_F=0.4\,$eV. The rate is given as a function of position along the dashed-line excursions shown in the upper insets. Plot (b) is obtained for a distance of 10\,nm from the graphene plane. The rate is calculated at the peak position of the lowest-order $m=1$ and $m=0$ resonances, respectively (see Fig.\ 4 of the main paper).} \label{FigSI8}
\end{figure}

In the main paper, we have discussed the decay rate for an emitter situated at a fixed point along the axis of self-standing circular graphene disks. Here we consider the variation of the peak decay rate as a function of position of the emitter. Figure\ \ref{FigSI8}a shows the variation of the rate with distance to the graphene along the axis of the disk. The rate decays with distance $z$ roughly as $\exp(-4\pi z/\lambda_{\rm sp})$ (i.e., it follows the same exponential attenuation as in homogeneous graphene), where the plasmon wavelength $\lambda_{\rm sp}$ is 290\,nm for the $m=1$ mode and 94\,nm for the $m=0$ mode. The variation along parallel displacements (Fig.\ \ref{FigSI8}b) is less trivial, but it qualitatively follows the near-field intensities shown in Fig.\ 4c,d of the main paper.

\begin{figure}
\centerline{\includegraphics*[width=17cm]{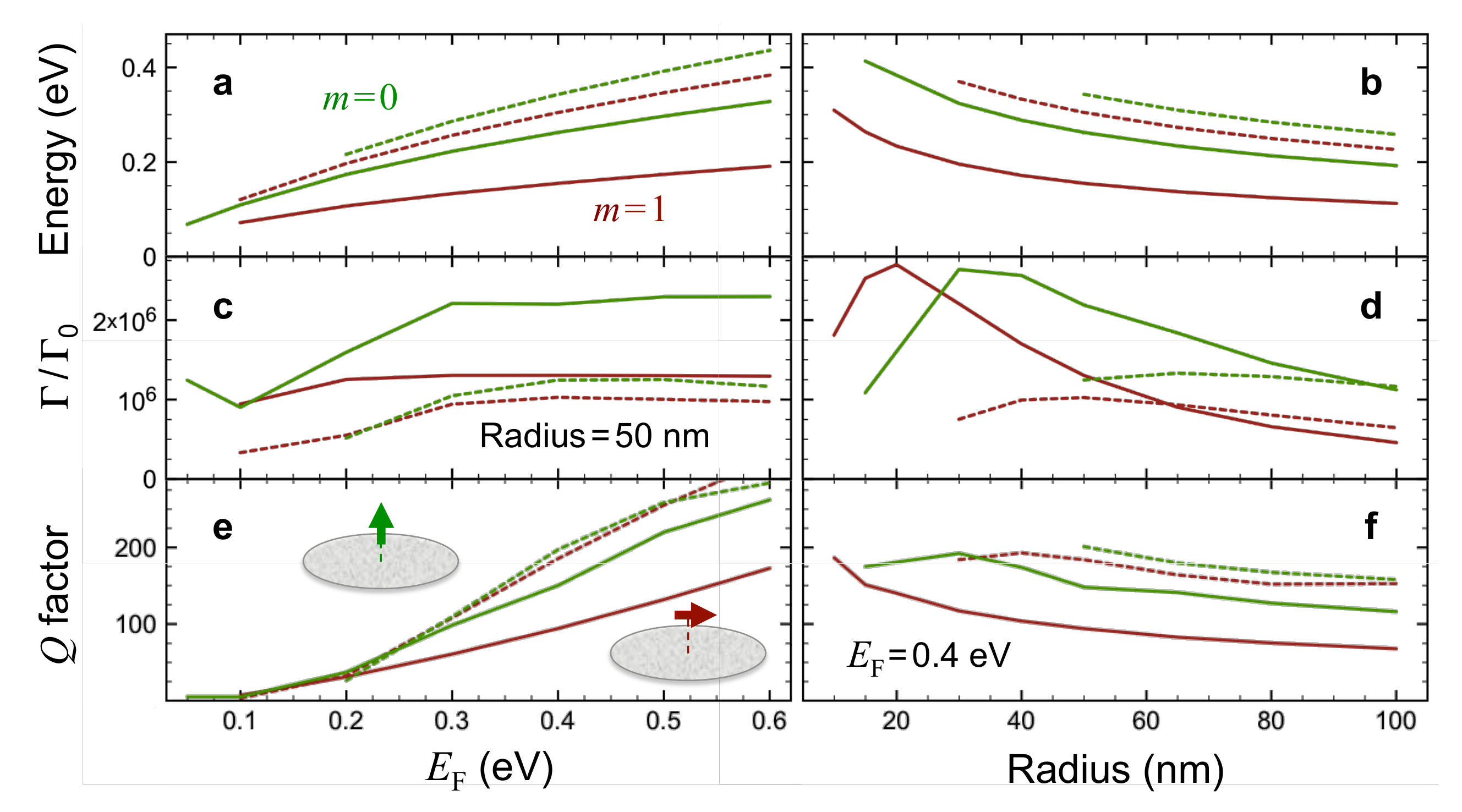}}
\caption{{\bf Plasmons in graphene nanodisks: Size and doping dependence.} {\bf (a,b)} Fermi-energy and disk-radius dependence of the SP energy in doped graphene disks for the first- (solid curves) and second-order (dashed curves) modes with either $m=0$ (green curves) or $m=1$ (red curves) azimuthal symmetries. The two lowest-order modes are shown for each symmetry. {\bf (c,d)} Decay rate $\Gamma$ of an emitter located 10\,nm above the center of the disk, normalized to the rate in free space $\Gamma_0$, and calculated at the energies of the SPs in (a,b). The emitter is polarized parallel (perpendicular) to the disk in the red (green) curves. {\bf (e,f)} Quality factor of the resonances considered in (a,b).} \label{FigSI9}
\end{figure}

\begin{figure}
\centerline{\includegraphics*[width=10cm]{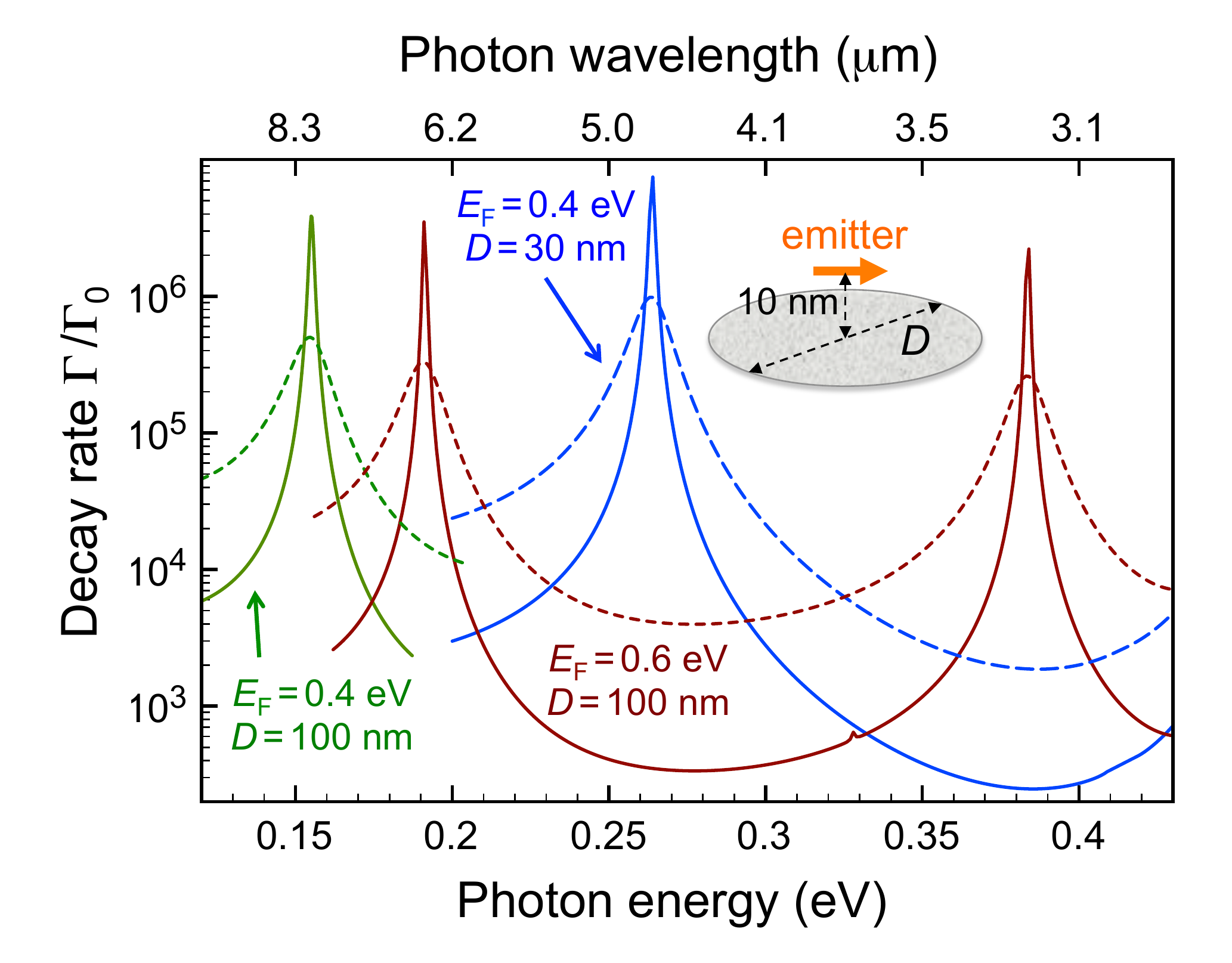}}
\caption{{\bf Plasmons in graphene nanodisks: Relaxation dependence.} Decay rate of an emitter situated 10\,nm above the center of a doped graphene disk for two different disk diameters $D$ and two values of the Fermi energy $E_F$, as shown by text labels. The emitter is polarized parallel to the disk. Dashed curves are calculated with $\tau\sim0.5\times10^{-13}\,$s. Solid curves are obtained with $\tau=4\times10^{-13}\,$s for $E_F=0.4\,$eV and $\tau=6\times10^{-13}\,$s for $E_F=0.6\,$eV.} \label{FigSI10}
\end{figure}

\subsection{Plasmons in graphene nanodisks: Size, doping, and relaxation dependence}

Figure\ \ref{FigSI9} summarizes the $E_F$ and disk-size dependence of SPs in graphene nanodisks. The evolution of the plasmon energy is inherited from the $\omega_p\propto\sqrt{E_F/\lambda_{\rm sp}}$ scaling in homogeneous graphene, so that it increases with $\sqrt{E_F}$ (Fig.\ \ref{FigSI9}a) and decreases with the inverse of the square root of the radius (Fig.\ \ref{FigSI9}b).

The peak decay rate has a weak dependence on both $E_F$ and disk size (Fig.\ \ref{FigSI9}c,d), essentially reflecting the increase in SP lifetime when its energy is positioned close to $E_F$. Maximum rates $\sim10^6\,\Gamma_0$ are consistently obtained near these disks at the distance $z=10\,$nm chosen in the calculations.

The quality factor of the SP resonances $Q$ (Fig.\ \ref{FigSI9}e,f, extracted from the peak frequency divided by the FWHM) shows a strong increase with $E_F$, in agreement with the longer propagation distance observed in homogeneous graphene at higher doping. Our reported values of $Q>100$ are consistent with the moderate relaxation time employed in the calculations, $\tau\approx10^{-13}\,$s (i.e., $Q\sim\omega_p\tau$). Actually, the values of $Q$ reported in Fig.\ \ref{FigSI9}e,f are qualitatively well predicted by this formula using the energies of Fig.\ \ref{FigSI9}a,b as input. At the same time, much higher values of the mobility $\mu$ have been reported\cite{BSJ08}, which should lead to larger $Q$'s and peak rates in direct proportion to $\mu$, at least below the optical phonon frequency.

The role of plasmons in the relaxation of graphene is not yet well understood, although careful analysis \cite{JBS09} reveals that their effect can be incorporated through an effective $\tau\sim0.5\times10^{-13}\,$s. We show in Fig.\ \ref{FigSI10} results for the decay rate near nanodisks obtained with this effective value of $\tau$ (dashed curves), compared to calculations obtained with $\tau=\mu E_F/ev_F^2$ and a mobility $\mu=10,000\,$cm$^2/$Vs (solid curves, taken from Fig.\ 4 of the main paper). The shorter relaxation time due to phonons produces a decrease in both the peak rate and $Q$ by a factor of $\sim$5-9 in the spectra shown here, which increases linearly with $E_F$ and decreases smoothly with $\omega$.

\subsection{Field lines in near-field plots}

The near-electric-fields shown in the main paper for ribbons and disks are obtained close to a non-degenerate resonance at $\omega=\omega_0$, so that they take the form $\Eb^{\rm ext}+\Fb/(\omega_0-\omega-i\gamma/2)$, where $\gamma$ is the plasmon relaxation rate. Since the ribbon width is much smaller than the light wavelength, the near field must be almost electrostatic, and therefore, $\Eb$ and $\Fb$ are approximately real vectors. Interestingly, the on-resonance induced field $i2\Fb/\gamma$ is almost imaginary, in good agreement with our numerical simulations. Using a real transition dipole, the induced fields plotted in the main paper are more than 99\% imaginary, and this actually allows us to extract the field lines that are shown there (i.e., field lines corresponding to a nearly real, electrostatic electric field).

\subsection{Polarizability of the combined SP-emitter system in the Jaynes-Cummings model}

The interaction with the external field $E(t)$ can be written \[H_{\rm ext}=-(P_p+P_0)\,E(t),\] which involves the plasmon and emitter dipole operators $P_p=d_pa+d_p^*a^+$ and $P_0=d_0\sigma+d_0^*\sigma^+$, where $d_p$ and $d_0$ are their respective transition dipoles.

We derive the polarizability $\alpha$ from the first-order-perturbation-theory steady-state solution of the model Hamiltonian for a faint external field of the form $E(t)=2{\rm Re}\{E_0e^{-i\omega t}\}$. The polarizability is defined through the relation $p(t)=2{\rm Re}\{\alpha(\omega)E_0e^{-i\omega t}\}$, obtained from the expected value of the combined induced dipole, $p=\langle P_p+P_0\rangle$. After some algebra, we find
\begin{equation}
\alpha(\omega)=\alpha_0(\omega)+\alpha_0^*(-\omega), \nonumber
\end{equation}
where
\begin{equation}
\alpha_0(\omega)=\frac{(\omega_0-\omega-i\Gamma_0/2)|d_p|^2+(\omega_p-\omega-i\kappa/2)|d_0|^2-2g{\rm Im}\{d_p d_0^*\}}{(\omega_0-\omega-i\Gamma_0/2)(\omega_p-\omega-i\kappa/2)-g^2}.
\nonumber
\end{equation}

In the calculations presented in Fig.\ 5c of the main paper we assume $d_0$ to be negligible compared to $d_p$. Furthermore, the model parameters $\omega_p$, $\Gamma$, and $Q$ are extracted from the second $m=1$ mode of Fig.\ \ref{FigSI9}a,c,e (red dashed curves) following the procedure described in the main paper. 

\bibliographystyle{achemso}
\bibliography{refs}

\end{document}